\documentclass[letterpaper,american,prx,twocolumn,nofootinbib,amsmath,amssymb,showpacs,superscriptaddress]{revtex4-1}
\usepackage[T1]{fontenc}
\usepackage{babel}
\usepackage{prettyref}
\usepackage{float}
\usepackage{amstext}
\usepackage{graphicx}
\usepackage[unicode=true,
 bookmarks=true,bookmarksnumbered=false,bookmarksopen=false,
 breaklinks=false,pdfborder={0 0 1},backref=false,colorlinks=false]
 {hyperref}
\hypersetup{
 pdfauthor={Dahmen, Bos, Helias},
 pdfsubject={Manuscript for PRX},
 colorlinks=true,linkcolor=black,citecolor=black,urlcolor=black,filecolor=black}
\usepackage{breakurl}

\makeatletter

\newrefformat{cap}{\hyperref[#1]{Figure~\ref{#1}}}
\newrefformat{fig}{\hyperref[#1]{Figure~\ref{#1}}}
\newrefformat{tab}{\hyperref[#1]{Table ~\ref{#1}}}
\newrefformat{sec}{\hyperref[#1]{Section~\ref{#1}}}
\newrefformat{sub}{\hyperref[#1]{Section~\ref{#1}}}
\newrefformat{cha}{\hyperref[#1]{Chapter~\ref{#1}}}

\usepackage{MnSymbol}  

\makeatother

\begin{document}
\global\long\def\ms{\:\mathrm{ms}}
\global\long\def\M{\mathbf{M}}
\global\long\def\I{\mathbf{I}}
\global\long\def\Var{\mathrm{Var}}
\global\long\def\E{\mathrm{E}}
\global\long\def\D{\mathbf{D}}
\global\long\def\a{\mathbf{a}}
\global\long\def\w{\mathbf{w}}
\global\long\def\v{\mathbf{v}}
\global\long\def\erfc{\mathrm{erfc}}
\global\long\def\n{\mathbf{n}}
\global\long\def\q{\mathbf{q}}
\global\long\def\diag{\mathrm{diag}}
\global\long\def\x{\mathbf{x}}
\global\long\def\t{\mathbf{t}}

\title{Correlated fluctuations in strongly-coupled binary networks beyond
equilibrium}

\author{David Dahmen}

\affiliation{Institute of Neuroscience and Medicine (INM-6) and Institute for
Advanced Simulation (IAS-6) and JARA BRAIN Institute I, Jülich Research
Centre, 52425 Jülich, Germany}

\author{Hannah Bos}

\affiliation{Institute of Neuroscience and Medicine (INM-6) and Institute for
Advanced Simulation (IAS-6) and JARA BRAIN Institute I, Jülich Research
Centre, 52425 Jülich, Germany}

\author{Moritz Helias}

\affiliation{Institute of Neuroscience and Medicine (INM-6) and Institute for
Advanced Simulation (IAS-6) and JARA BRAIN Institute I, Jülich Research
Centre, 52425 Jülich, Germany}

\affiliation{Department of Physics, Faculty 1, RWTH Aachen University, 52074 Aachen,
Germany}

\date{\today}

\pacs{64.60.De, 75.10.Nr, 87.19.lj, 05.70.Ln }
\begin{abstract}

Randomly coupled Ising spins constitute the classical model of collective
phenomena in disordered systems, with applications covering ferromagnetism,
combinatorial optimization, protein folding, stock market dynamics,
and social dynamics. The phase diagram of these systems is obtained
in the thermodynamic limit by averaging over the quenched randomness
of the couplings. However, many applications require the statistics
of activity for a single realization of the possibly asymmetric couplings
in finite-sized networks. Examples include reconstruction of couplings
from the observed dynamics, learning in the central nervous system
by correlation-sensitive synaptic plasticity, and representation of
probability distributions for sampling-based inference. The systematic
cumulant expansion for kinetic binary (Ising) threshold units with
strong, random and asymmetric couplings presented here goes beyond
mean-field theory and is applicable outside thermodynamic equilibrium;
a system of approximate non-linear equations predicts average activities
and pairwise covariances in quantitative agreement with full simulations
down to hundreds of units. The linearized theory yields an expansion
of the correlation- and response functions in collective eigenmodes,
leads to an efficient algorithm solving the inverse problem, and shows
that correlations are invariant under scaling of the interaction strengths.

\end{abstract}
\maketitle

\section{Introduction}

Understanding collective phenomena arising from the interactions
in a many-body system is the challenging subject of many-particle
physics. The characterization of the emerging states typically rests
on the quantification of correlations \citep{Binney92}. Among the
simplest classical models is the Ising (binary) model \citep{Ising25_253},
or Glauber dynamics \citep{Glauber63_294} in its kinetic formulation.
While for symmetric random couplings \citep{Sherrington75_1792} these
systems are within the realm of equilibrium statistical mechanics
\citep{Thouless77_593}, the asymmetric kinetic Ising model \citep{Hertz86_151,Derrida87_721,Crisanti87_4922},
even in the stationary state, does not reach thermodynamic equilibrium.
In their Markovian formulation \citep{Kelly79} these processes are
studied in the field of non-equilibrium stochastic thermodynamics
\citep[for review see][chapter 6]{Seifert12_126001}. The methods
derived in these fields have proven useful in a variety of disciplines,
including computer science, biology, artificial intelligence, social
sciences and economics \citep[see e.g.][and references therein]{Mezard87,Nishimori01_01,Barrat08,Castellano09_591,Stein13,Sornette14_062001}. 

Neuronal networks of the central nervous system are prominent examples
of non-equilibrium systems due to Dale's principle \citep{Eccles54_524,Li99_59},
which states that a neuron either excites or inhibits all its targets.
The correlations between the activities of nerve cells are functionally
important \citep{Cohen11_811}: they influence the representation
of information in population signals depending on the readout in an
either detrimental \citep{Zohary94_140} or beneficial \citep{Shamir01_277}
manner, their time-dependent modulations are linked to behavior \citep{Kilavik09_12653},
and they determine the influence of neuronal activity on synaptic
plasticity \citep{Markram97a,Bi98,Morrison08_459}, the biophysical
substrate underlying learning. In case of binary units (Ising spins),
knowledge of pairwise covariances, moreover, proves useful for sampling-based
inference \citep[reviewed in][]{Fiser10_119} as they constitute the
next order in the systematic expansion of the joint probability distribution
beyond independence \citep[cf.][ eq. (22)]{Glauber63_294}.

Dynamic mean-field theory \citep{Sompolinsky82_6860,Sompolinsky88_259,Aljadeff15_088101,Kadmon15_041030}
in the $N\to\infty$ limit effectively reduces the many body problem
of a network comprised of a large number of units to a single particle
interacting with a self-consistently determined field, but neglects
the cross-covariances of activities between units. \citet{Ginzburg94}
introduced pairwise correlations to the description of weakly coupled
systems and \citet{Renart10_587} extended the analysis to strongly
coupled units in the large $N$ limit. Both approaches are, however,
limited to averaged pairwise correlations. Similarly, approximate
master equations for binary-state dynamics on complex networks, as
recently discussed in context of the pair approximation by \citet{Gleeson11_068701,Gleeson13_021004},
are restricted to global dynamics in infinite networks. 

In the current work, we derive a systematic cumulant expansion yielding
an analytical description of correlations between the activities of
individual pairs of units, which goes beyond averaged pairwise correlations
\citep{Ginzburg94,Buice09_377,Renart10_587} and is applicable to
a single realization of an asymmetric (directed) network. The framework
links the coupling structure to the emerging correlated activity and
describes the first and second order statistics for single units and
pairs of units in a large, but finite, network of possibly strongly
interacting elements. The analytical expressions predict a distribution
of pairwise correlations with a small mean but large standard deviation,
as observed experimentally \citep{Ecker10} in neural tissue and yet
not explained theoretically. Moreover, the framework can be employed
to infer the couplings between the units from the observed correlated
activity, also termed the inverse problem \citep{Aertsen90,Zheng11_041135,Mezard11_L07001,Tyrcha13_1301}.

In particular, we show that already a Gaussian truncation of the presented
cumulant hierarchy yields good predictions for individual mean activities
and pairwise correlations. Taking into account the subset of third
order cumulants that for binary variables can be expressed by lower
order ones, improves the prediction significantly. This finding demonstrates
that the second order statistics suffices to capture the major features
of the collective dynamics arising in random binary networks, even
when the coupling is strong. Our approach consistently takes into
account the network-generated fluctuations in the marginal statistics
of the input to each unit in a similar spirit as the seminal work
by \citet{Vreeswijk96}. This approach exposes peculiar features of
networks of binary threshold units. For a fixed number of incoming
connections to each unit, mean-field theory predicts their averaged
activities to be identical. We here show how distributed mean activities
arise solely from the correlations emerging in the network. Units
with a hard activation threshold render covariances independent of
synaptic amplitudes $J$: if all incoming connections as well as the
threshold of a unit are changed proportionally, the mean activity
and correlations are maintained. Hence, scaling the threshold appropriately,
it is impossible to increase the influence of one unit on another
by larger coupling amplitudes. This finding questions the customary
division into strong ($J\propto N^{-\frac{1}{2}}$) and weak coupling
($J\propto N^{-1}$) in such systems. We here show in addition that
a network with hard threshold units implements the strongest possible
coupling between stochastic binary units.

The independence of covariances with respect to coupling strengths
implies that the latter cannot be reconstructed from the knowledge
of the activity alone. However, the amplitude and slope of covariance
functions at zero time lag together uniquely determine a linearized
effective coupling strength between units. A linear approximation
of the dynamics of fluctuations around the stationary state leads
to a modified Lyapunov equation. Decomposing the fluctuations into
characteristic eigenmodes of the network shows that, to linear order,
the kinetic binary network is equivalent to a system of coupled Ornstein-Uhlenbeck
processes \citep{Uhlenbeck30,Risken96}. The linear description further
allows for the decomposition of the response to external stimulations
into eigenmodes of the system.

\section{First and second moments of the joint probability distribution\label{sec:moment_equations}}

We here consider the classical network model of stochastic binary
units, and denote the activity of unit $k$ as $n_{k}(t)$ being either
$0$ or $1$, where $1$ indicates activity and $0$ inactivity \citep[see e.g.][]{Hertz91,Ginzburg94,Vreeswijk96,Buice09_377,Renart10_587,Gleeson11_068701,Gleeson13_021004},
following the notation of \citet{Buice09_377}. Since we may interpret
a unit as representing an individual neuron, we use the terms ``unit''
and ``neuron'' interchangeably. The state of the network of $N$
such units is described by a binary vector $\n=(n_{1},\ldots,n_{N})\in\{0,1\}^{N}$.
The model shows transitions at random points in time between the two
states $0$ and $1$ controlled by transition probabilities. Using
asynchronous update \citep{PDP86a}, in each infinitesimal interval
$[t,t+\delta t)$ each unit in the network has the probability $\frac{1}{\tau}\delta t$
to be chosen for update \citep{Hopfield82}, where $\tau$ is the
time constant of the dynamics. An equivalent implementation draws
the time points of update independently for all units. For a particular
unit, the sequence of update points has exponentially distributed
intervals with mean duration $\tau$, i.e. the update times form a
Poisson process with rate $\tau^{-1}$. We employ the latter implementation
in the globally time-driven \citep{Hanuschkin10_113} spiking simulator
NEST \citep{Gewaltig_07_11204}, and use a discrete time resolution
$\delta t=0.1\ms$ for the intervals. The stochastic update constitutes
a source of noise in the system. Given the $k$-th neuron is selected
for update, the probability to end in the up-state ($n_{k}=1$) is
determined by the activation function $F_{k}(\n)$ which depends on
the activity $\n$ of all connected units. The probability to end
in the down state ($n_{k}=0$) is given by $1-F_{k}(\n)$. 

The stochastic system at time $t$ is completely determined by its
probability distribution $p(\n,t)$. The time evolution of the two
point distribution $p(\n,t,\q,s)$ (describing the probability that
the system was in state $\q$ at time $s$ and is in state $\n$ at
time $t$) obeys - due to the Markov property - the same master equation
(see \prettyref{eq:balancing_flux} in \prettyref{sub:Derivation-of-moment-equations})
as $p(\n,t)$. Using the definitions of the first moment and the equal
time as well as the two time point second moments

\begin{eqnarray}
\langle n_{k}(t)\rangle & := & \sum_{\n}p(\n,t)\,n_{k}\label{eq:def_moments}\\
\langle n_{k}(t)n_{l}(t)\rangle & := & \sum_{\n}p(\n,t)\,n_{k}n_{l}\nonumber \\
\langle n_{k}(t)n_{l}(s)\rangle & := & \sum_{\n,\q}p(\n,t,\q,s)\,n_{k}q_{l},\nonumber 
\end{eqnarray}
one obtains a set of differential equations for the first and second
moments \citep{Ginzburg94,Buice09_377,Renart10_587,Helias14}, which
read (for completeness, in the Appendix we included their derivation
in \prettyref{eq:second_moment_diffeq})

\begin{widetext}

\begin{eqnarray}
\tau\frac{\partial}{\partial t}\langle n_{k}(t)\rangle & = & -\langle n_{k}(t)\rangle+\langle F_{k}(\n(t))\rangle\nonumber \\
k\neq l\qquad\tau\frac{\partial}{\partial t}\langle n_{k}(t)n_{l}(t)\rangle & = & -2\langle n_{k}(t)\,n_{l}(t)\rangle+\langle F_{k}(\n(t))\,n_{l}(t)\rangle+\langle F_{l}(\n(t))\,n_{k}(t)\rangle\label{eq:1st_2nd_moment}\\
t>s\qquad\tau\frac{\partial}{\partial t}\langle n_{k}(t)n_{l}(s)\rangle & = & -\langle n_{k}(t)\,n_{l}(s)\rangle+\langle F_{k}(\n(t))\,n_{l}(s)\rangle,\nonumber 
\end{eqnarray}
\end{widetext} where the expectation value $\langle\rangle$ defined
in \eqref{eq:def_moments} can be interpreted as an average over realizations
of the random dynamics. Note that the second line does not hold for
$k=l$, but becomes the first line due to $\langle n_{k}^{2}\rangle=\langle n_{k}\rangle=:m_{k}$.
The third line becomes the second line for $t\to s$. These equations
are identical to eqs.\ (3.4)-(3.7) in \citep{Ginzburg94}, to eqs.\ (3.12)
and (3.13) in \citep{Buice09_377}, and to eqs.\ (18)-(22) in \citep[Supplementary material]{Renart10_587}.
These previous works have so far considered the second order statistics
averaged over many pairs of neurons, i.e. $\langle n_{\alpha}n_{\beta}\rangle=\frac{1}{N_{\alpha}N_{\beta}}\,\sum_{i\in\alpha\,j\in\beta}\langle n_{i}n_{j}\rangle$.
These averages are closely related to the population-averaged activities
$n_{\alpha}(t)=\frac{1}{N_{\alpha}}\sum_{i\in\alpha}n_{i}(t)$ and
can therefore be treated by population-averaging mean-field methods.
Here, we go beyond the population-averaged level and consider the
second order statistics of individual pairs. Initially, we are interested
in the stationary statistics of the mean activities of individual
units and their zero time lag pairwise covariances $c_{kl}=\langle n_{k}(t)n_{l}(t)\rangle-\langle n_{k}(t)\rangle\langle n_{l}(t)\rangle$,
which can be deduced from \eqref{eq:1st_2nd_moment} 

\begin{eqnarray}
m_{k} & = & \langle F_{k}(\mathbf{n})\rangle\nonumber \\
c_{kl} & = & \begin{cases}
\frac{1}{2}\langle F_{k}(\mathbf{n})n_{l}\rangle+\frac{1}{2}\langle F_{l}(\mathbf{n})n_{k}\rangle-m_{k}m_{l} & \quad k\neq l\\
m_{k}(1-m_{k}) & \quad k=l
\end{cases}.\label{eq:stationary_statistics}
\end{eqnarray}

\section{Gaussian approximation\label{sec:Gaussian-approximation}}

Following \citep{Vreeswijk96,Renart10_587}, we now assume a particular
form for the activation function and couplings between neurons given
by

\begin{eqnarray}
F_{k}(\n) & = & f(h_{k}(\n))\label{eq:gain_function}\\
h_{k} & = & \sum_{i=1}^{N}J_{ki}n_{i},\nonumber 
\end{eqnarray}
where the gain function $F_{k}(\n)=f(h_{k}(\n))$ depends on the state
of the network only via the summed and weighted activity $h_{k}$,
which is a scalar. Here $J_{ki}$ denotes the matrix of synaptic couplings
from neuron $i$ to neuron $k$. For concreteness, when comparing
the analytical predictions to numerical results, we choose $f(h_{k})=H\left(h_{k}-\theta\right)$,
where $H(x)=1\quad\text{for }x\ge0,\quad0\quad\text{else}$ is the
Heaviside function. Intermediate results also hold for arbitrary gain
functions $f$. If the distribution of $h_{k}$ would be known, the
expectation value $\langle F_{k}\rangle$ could directly be calculated.

We aim at a self-consistent equation for the first and second moments
of the activity variables \eqref{eq:1st_2nd_moment}. Van Vreeswijk
et al. \citep{Vreeswijk96} and \citet{Renart10_587} invoke the central
limit theorem to justify the assumption that the local field $h_{k}$
typically follows a Gaussian distribution $\mathcal{N}(\mu_{k},\sigma_{k}^{2})$
with cumulants $\mu_{k}$ and $\sigma_{k}^{2}$. These cumulants are
related to cumulants of the activity variables $\n$ via
\begin{eqnarray}
\mu_{k} & = & \langle h_{k}\rangle=\sum_{i}J_{ki}\langle n_{i}\rangle=(\mathbf{J}\,\mathbf{m})_{i}\label{eq:mu_sigma}\\
\sigma_{k}^{2} & = & \langle h_{k}^{2}\rangle-\mu_{k}^{2}=\sum_{i,j}J_{ki}J_{kj}\left(\langle n_{i}n_{j}\rangle-\langle n_{i}\rangle\langle n_{j}\rangle\right)\nonumber \\
 & = & \left(\mathbf{J}\,\mathbf{C}\,\mathbf{J}^{T}\right)_{kk},\nonumber 
\end{eqnarray}
with the covariance matrix $c_{ij}=\langle n_{i}n_{j}\rangle-\langle n_{i}\rangle\langle n_{j}\rangle$,
which contains $c_{ii}=\langle n_{i}\rangle(1-\langle n_{i}\rangle)$
on the diagonal and the vector of mean activities $m_{i}=\langle n_{i}\rangle$.
In the seminal work by \citep{Vreeswijk98}, the influence of the
cross-covariances on the variance in the input to a neuron was neglected.
Instead, only the dominant contribution, given by the variances $c_{ii}$
of the single neurons on the diagonal, were taken into account, leading
to the expression $\sigma_{k}^{2}=\sum_{i}J_{ki}^{2}m_{i}(1-m_{i})$.
The additional dependence of $\sigma_{k}^{2}$ on the off-diagonal
elements \eqref{eq:mu_sigma} is important for the distribution of
mean activities, as we will show in the following. We note that due
to the marginal binary statistics of $n_{i}$ it follows that $\langle n_{i}^{2}\rangle=\langle n_{i}\rangle$,
illustrating that the second moment is uniquely determined by the
first moment. We will exploit this property in the subsequent sections
by expressing a subset of higher order cumulants in terms of lower
order ones. Using the central-limit theorem argument, as outlined
above, leads to 
\begin{eqnarray}
m_{k}=\langle F_{k}(\n)\rangle & \simeq & \int_{-\infty}^{\infty}H(x-\theta)\,\mathcal{N}(\mu_{k},\sigma_{k}^{2},x)\,dx\label{eq:mean_erfc}\\
 & = & \frac{1}{2}\erfc\left(-\frac{\mu_{k}-\theta}{\sqrt{2}\sigma_{k}}\right).\nonumber 
\end{eqnarray}
To enable the extension to further corrections, we aim at a more systematic
calculation of expectation values containing the gain function. Using
the Fourier representation $f(x)=\frac{1}{2\pi}\int_{-\infty}^{\infty}d\omega\,\hat{f}(\omega)\,\exp\left(i\omega\,x\right)$
of the gain function $f$ we obtain 
\begin{eqnarray}
\langle F_{k}(\n)\rangle_{\n} & = & \frac{1}{2\pi}\int_{-\infty}^{\infty}d\omega\,\hat{f}(\omega)\left\langle \exp\left((i\omega)\sum_{j=1}^{N}J_{kj}n_{j}\right)\right\rangle _{\n}\label{eq:gain_expectation-1-1}\\
 & = & \frac{1}{2\pi}\int_{-\infty}^{\infty}d\omega\,\hat{f}(\omega)\left\langle \exp\left((i\omega)\,h_{k}(\n)\right)\right\rangle _{\n},\nonumber 
\end{eqnarray}
\begin{figure}[H]
\begin{centering}
\includegraphics{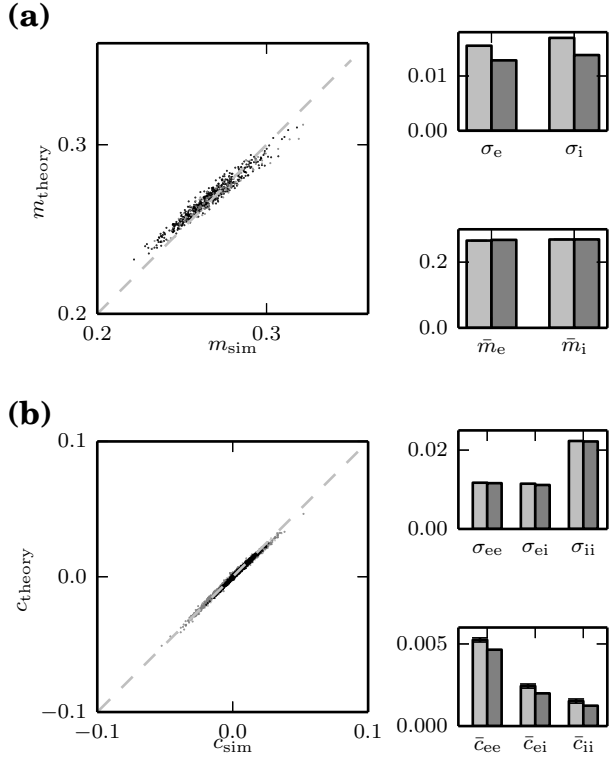}
\par\end{centering}

\caption{\textbf{Consistent Gaussian approximation for the mean activity and
covariance matrix in a random network of binary neurons. (a) }Theoretical
prediction of mean activities \eqref{eq:lin_sol} versus simulation
results (black: excitatory, gray: inhibitory). The diagonal is indicated
by the dashed line. Lower inset: mean activity averaged over excitatory
and inhibitory neurons (light gray: simulation, darker gray: theory).
Black error bars show standard error of the mean obtained from $20$
simulations (not visible, below line width). Upper inset: width of
distribution of mean activities.\textbf{ (b)} Theoretical prediction
\eqref{eq:lin_sol} versus simulated covariance. Only cross-covariances
are shown, as auto-covariances follow trivially by \eqref{eq:stationary_statistics}
from the mean activities. Lower inset: mean covariances averaged over
excitatory-excitatory, excitatory-inhibitory and over inhibitory-inhibitory
neuron pairs (light gray: simulation, darker gray: theory). Error
bars show standard error of the mean obtained from $20$ simulations.
Upper inset: width of the distributions of covariances.\textbf{ }Network
parameters: number of excitatory ($N_{E}=500$) and inhibitory neurons
($N_{I}=125$), threshold $\theta=-5.5$, excitatory coupling strength
$J_{ij}=J=1,\,\forall j\in E$, inhibitory coupling strength $J_{ij}=-6\,J,\,\forall j\in I$,
fixed in-degree homogeneous random network with connection probability
$p=0.2$. Results averaged over $20$ simulations with $T=2,000,000\protect\ms$
each. Theoretical predictions are obtained by damped fixed-point iteration
\prettyref{eq:damped_fixed_point}.\label{fig:covariance_lin}}
\end{figure}
where we inserted the definition of $h_{k}$ \eqref{eq:gain_function}
in the second line. Defining the vector $\mathbf{t}$ with elements
$t_{j}=i\omega J_{kj}$, the expectation value in the first line of
the previous expression has the form $\langle\exp(\t\cdot\n)\rangle_{\n}$,
which is the definition the characteristic function or moment generating
function 
\begin{eqnarray}
\varphi_{n}(\mathbf{t}) & \equiv & \langle e^{\mathbf{t}\cdot\mathbf{n}}\rangle_{\n}\label{eq:moment_generator-1}\\
 & = & \sum_{\n\in\{0,1\}^{N}}p(\n)\,e^{\mathbf{t}\cdot\n}\nonumber \\
 & = & \exp\left(\Phi_{\n}(\mathbf{t})\right)\nonumber 
\end{eqnarray}
of the joint distribution of the binary variables $p(\n)$ \citep[p. 32]{Gardiner85},
which can also be expressed by the cumulant generating function $\Phi_{\n}(\mathbf{t})$.
Note that for compactness of notation we omitted the index $j$ for
$\mathbf{t}$. The Taylor expansion of $\Phi_{\n}(\mathbf{t})$ explicitly
introduces a cumulant hierarchy showing that cumulants of activity
variables $n_{j}$ at all orders contribute to $\langle F_{k}(\n)\rangle.$
If we neglect cumulants higher than order two, thus effectively approximating
the binary states as correlated Gaussian variables, we use the corresponding
quadratic cumulant generating function 
\begin{eqnarray}
\Phi_{\n}(\mathbf{t}) & = & \sum_{i}m_{i}t_{i}+\frac{1}{2}\sum_{i,j}c_{ij}\,t_{i}t_{j}\label{eq:quadratic_cumulant}
\end{eqnarray}
and obtain
\begin{eqnarray}
\langle F_{k}(\n)\rangle_{\n} & \simeq & \frac{1}{2\pi}\int_{-\infty}^{\infty}d\omega\,\hat{f}(\omega)\,\exp\left(\sum_{i}m_{i}t_{i}+\frac{1}{2}\sum_{i,j}c_{ij}\,t_{i}t_{j}\right)\nonumber \\
 & = & \frac{1}{2\pi}\int_{-\infty}^{\infty}d\omega\,\hat{f}(\omega)\,\exp\left(\mu_{k}i\omega+\frac{1}{2}\sigma_{k}^{2}(i\omega)^{2}\right).\label{eq:F_i_Fourier-1}
\end{eqnarray}
Here we identified the sums in the penultimate line as the mean
and variance of the input field \eqref{eq:mu_sigma}. On the other
hand in the second line of \eqref{eq:gain_expectation-1-1} we identify
the moment generating function $\varphi_{h_{k}}(i\omega)\equiv\left\langle \exp\left((i\omega)\,h_{k}(\n)\right)\right\rangle _{\n}$
of the input field $h_{k}$. From \eqref{eq:F_i_Fourier-1} we obtain
its corresponding cumulant generating function as $\Phi_{h_{k}}(i\omega)=\mu_{k}i\omega+\frac{1}{2}\sigma_{k}^{2}(i\omega)^{2}$.
We therefore conclude that $h_{k}\sim\mathcal{N}(\mu_{k},\sigma_{k}^{2})$
follows a Gaussian distribution. The mean activity then writes
\begin{eqnarray}
m_{k}=\langle F_{k}(\n)\rangle_{\n} & = & \langle f(h_{k})\rangle_{h_{k}\sim\mathcal{N}(\mu_{k},\sigma_{k}^{2})},\label{eq:general_mean_gaussian}
\end{eqnarray}
which for $f(x)=H(x-\theta)$ is identical to \eqref{eq:mean_erfc}.
The procedure hence reproduces the known result for the Heaviside
gain function \citep[supplementary material, Sections 2.2, 2.3]{VanVreeswijk98_1321,Renart10_587},
but additionally yields a generalization for smooth gain functions
$f$ that takes into account the fluctuations of the synaptic input,
which is in line with the treatment in \citep[ eq. (27)]{Grytskyy13_258}.

 More importantly, this systematic calculation reveals the assumption
that is underlying the Gaussian approximation for the input field,
namely that cumulants higher than order two are ignored on the level
of individual neuronal activities.

\begin{figure}
\begin{centering}
\includegraphics{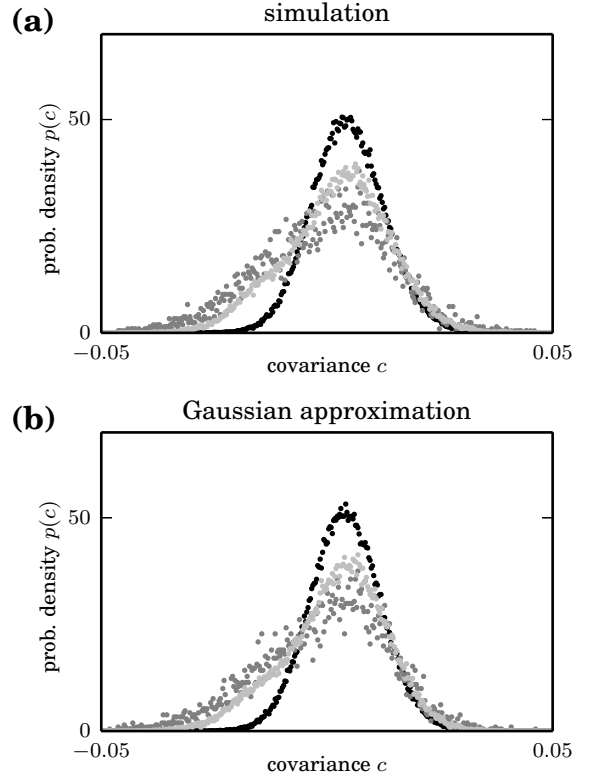}
\par\end{centering}

\caption{\textbf{Distribution of covariances in the Gaussian approximation
for a random network of binary neurons. (a)} Distribution of cross-covariance
between excitatory neurons (black), inhibitory neurons (gray) and
between one inhibitory, one excitatory neuron (light gray) from simulation.
\textbf{(b)} As (a), but showing theoretical result for the Gaussian
approximation \eqref{eq:lin_sol}. Parameters as in \prettyref{fig:covariance_lin}.\label{fig:covariance_lin-1}}
\end{figure}

To obtain an expression for the zero time lag covariance, we start
from \eqref{eq:stationary_statistics}. We hence need to evaluate
expressions of the form $\langle F_{k}(\n)n_{l}\rangle$ which do
not factorize trivially, since the value of $n_{l}$ may have an influence
on neuron $k$ through the gain function $F_{k}(\mathbf{n})$. However,
along the same lines as above, we can derive $\langle F_{k}(\n)n_{l}\rangle$
using the identical approximation of $\langle\exp\left(\mathbf{n}\cdot\mathbf{t}\right)\rangle$
by first noting

\begin{eqnarray*}
\langle F_{k}(\n)n_{l}\rangle & = & \frac{1}{2\pi}\int_{-\infty}^{\infty}d\omega\,\hat{f}(\omega)\left\langle \exp\left(i\omega\sum_{j=1}^{N}J_{kj}n_{j}\right)n_{l}\right\rangle \\
 & = & \frac{1}{2\pi}\int_{-\infty}^{\infty}d\omega\,\hat{f}(\omega)\,\frac{\partial}{\partial t_{l}}\langle\exp\left(\mathbf{n}\cdot\mathbf{t}\right)\rangle.
\end{eqnarray*}
Expressing again the characteristic function by the cumulant generating
function \eqref{eq:quadratic_cumulant} yields
\begin{align}
\frac{\partial}{\partial t_{l}}\exp\left(\Phi_{\n}(\mathbf{t})\right) & =\underbrace{\frac{\partial\Phi_{\n}(\t)}{\partial t_{l}}}_{=m_{l}+\sum_{j}J_{kj}c_{jl}(i\omega)\text{ by \eqref{eq:quadratic_cumulant}}}\,\exp\left(\Phi_{\n}(\mathbf{t})\right)\nonumber \\
 & =\left(m_{l}+\sum_{j}J_{kj}c_{jl}\frac{\partial}{\partial\mu_{k}}\right)\,\exp\left(\Phi_{h_{k}}(i\omega)\right),\label{eq:Gaussian_deriv_cumulant}
\end{align}
where we generated the factor $i\omega$ by a derivative $\partial_{\mu_{k}}$
in the last line. With \eqref{eq:general_mean_gaussian} this yields
\begin{eqnarray}
\langle F_{k}(\n)n_{l}\rangle & = & m_{l}\langle f(h_{k})\rangle_{h_{k}\sim\mathcal{N}(\mu_{k},\sigma_{k}^{2})}+S_{k}\left(JC\right)_{kl}\label{eq:lFnjr}\\
S_{k} & = & \frac{\partial}{\partial\mu_{k}}\langle f(h_{k})\rangle_{h_{k}\sim\mathcal{N}(\mu_{k},\sigma_{k}^{2})}\nonumber \\
 & \stackrel{\text{i.b.p.}}{=} & \langle f^{\prime}(h_{k})\rangle_{h_{k}\sim\mathcal{N}(\mu_{k},\sigma_{k}^{2})},\nonumber 
\end{eqnarray}
where we defined the local susceptibility $S_{k}$ that determines
the influence of an input to neuron $k$ onto its output at the time
point of its update. For a differentiable gain function $f$ the susceptibility
is equal to the slope $f^{\prime}$ averaged over the fluctuations
of $h_{k}$ \citep[compare also ][ eq. (27)]{Grytskyy13_258}, which
follows from integrating  the second line \eqref{eq:lFnjr} by parts
(i.b.p.). For the particular choice $f(x)=H(x-\theta)$, the susceptibility
has the form 
\begin{eqnarray}
S_{k} & := & \frac{1}{\sqrt{2\pi}\sigma_{k}}\,e^{-\frac{(\mu_{k}-\theta)}{2\sigma_{k}^{2}}^{2}},\label{eq:susceptibility-1}
\end{eqnarray}
which, by \eqref{eq:lFnjr}, is the strongest possible coupling for
a given input statistics. The self-consistent set of equations for
the first and second cumulants thus reads
\begin{eqnarray}
m_{k} & = & \frac{1}{2}\erfc\left(-\frac{\mu_{k}-\theta}{\sqrt{2}\sigma_{k}}\right)\nonumber \\
c_{kl} & = & \begin{cases}
\frac{1}{2}S_{k}\left(\mathbf{JC}\right)_{kl}+\frac{1}{2}S_{l}\left(\mathbf{JC}\right)_{lk} & \quad k\neq l\\
m_{k}(1-m_{k}) & \quad k=l
\end{cases}.\label{eq:lin_sol}
\end{eqnarray}
To obtain the joint solution of equations \prettyref{eq:lin_sol},
we use a damped fixed point iteration with the $n$-th iteration value
denoted as $(m_{k}^{(n)},c_{kl}^{(n)})$, which has the form
\begin{align}
m_{k}^{(n+1)} & \underset{\hphantom{\text{\ensuremath{{\scriptstyle \mbox{\ensuremath{k\neq l}}}}}}}{=}\rho\frac{1}{2}\erfc\left(-\frac{\mu_{k}^{(n)}-\theta}{\sqrt{2}\sigma_{k}^{(n)}}\right)+(1-\rho)m_{k}^{(n)}\nonumber \\
c_{kl}^{(n+1)} & \underset{\text{\ensuremath{{\scriptstyle \mbox{\ensuremath{k\neq l}}}}}}{=}\rho\frac{1}{2}\left(S_{k}^{(n)}\left(\mathbf{J}\mathbf{C}^{(n)}\right)_{kl}+S_{l}^{(n)}\left(\mathbf{J}\mathbf{C}^{(n)}\right)_{lk}\right)\nonumber \\
 & \underset{\hphantom{\text{\ensuremath{{\scriptstyle \mbox{\ensuremath{k\neq l}}}}}}}{+}(1-\rho)c_{kl}^{(n)}\nonumber \\
c_{kk}^{(n+1)} & \underset{\text{\ensuremath{{\scriptstyle \mbox{\ensuremath{k=l}}}}}}{=}m_{k}^{(n+1)}(1-m_{k}^{(n+1)}),\label{eq:damped_fixed_point}
\end{align}
where we used the damping factor $\rho=0.7$ and determined $\mu_{k}$
and $\sigma_{k}$ via \eqref{eq:mu_sigma} in each step. The iteration
continues until the summed absolute change of $m_{k}$ and $c_{kl}$
is smaller than $10^{-14}$. \prettyref{fig:covariance_lin} shows
that the theoretical prediction yields mean activities and covariances
in line with simulations of a random fixed-indegree network of excitatory
and inhibitory neurons. While cross-covariances are explained with
good accuracy, mean activities slightly, but systematically, deviate
from simulated results: The scatter plot, showing the mean activities
of individual neurons, has a slope below unity, indicating that the
width of the distribution is underestimated by the theory. Moreover,
we observe that the population-averaged covariances are slightly underestimated
by the theory. However, in summary, the theory in the Gaussian approximation
captures not only the mean and width of the covariance distribution
to a large extend but also its general shape as shown in \prettyref{fig:covariance_lin-1}.

The systematic calculation presented here shows that the Gaussian
assumption on the level of the individual statistics directly yields
the linearized result of \citep[supplementary material, Sections  2.3]{Renart10_587}.
Originally, the terms of the form $\langle F_{k}(\n)n_{l}\rangle$
in \eqref{eq:stationary_statistics} were obtained using relations
between conditioned and unconditioned probability distributions of
binary variables. After the Gaussian approximation, the resulting
non-linear equation was linearized and second order terms were discussed
to be small in the large $N$ limit. In contrast, the derivation here
shows that the linearization is not an additional assumption, but
appears naturally once the Gaussian approximation has been performed
on the level of individual variables. Note also, that no additional
assumption regarding the strength of the coupling is necessary.

\section{Third order cumulants\label{sec:Third-order-cumulants}}

\begin{figure}
\begin{centering}
\includegraphics{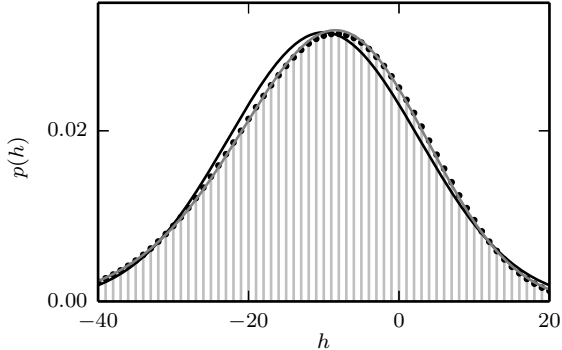}
\par\end{centering}

\caption{\textbf{Distribution of the summed input.} Circles show the distribution
assuming uncorrelated activity of the input neurons with constant
rate $m=0.2$, i.e. $p(h)=\sum_{k,l}B(pN_{E},m,k)B(pN_{I},m,l)\,\delta_{h,\:J_{E}\,k+J_{I}\,l}$
with binomial probabilities $B(N,p,k)$. Gaussian approximation $\mathcal{N}(\mu,\sigma^{2},h)$
(black) with same moments as summed binomial distribution $\mu=p\left(N_{E}J_{E}+N_{I}J_{I}\right)\,m$,
$\sigma^{2}=p\left(N_{E}J_{E}^{2}+N_{I}J_{I}^{2}\right)\,m(1-m)$.
Close-to-Gaussian approximation $p(h)\simeq\left(1+\frac{\kappa_{3}}{6}\,\partial_{\mu}^{3}\right)\mathcal{N}(\mu,\sigma^{2},h)$
(gray) taking into account the third cumulant $\kappa_{3}=p\left(N_{E}J_{E}^{3}+N_{I}J_{I}^{3}\right)\,(2m^{3}-3m^{2}+m)$
of $h$. Other parameters as in \prettyref{fig:covariance_lin}.\label{fig:close_to_Gaussian_approximation}}
\end{figure}

The derivation in the previous section is systematic in the sense
that all cumulants of order three and higher of the stochastic variables
$\n$ are consistently neglected. The original idea of a mean-field
description, however, seeks an approximation of the local field $h_{k}$
sensed by an individual unit $k$ rather than a truncation of the
cumulants of the individual variables $\n$ themselves. Since the
local field is by \eqref{eq:gain_function} a superposition of a large
number of weakly correlated contributions, the distribution of $h_{k}$
will be close to Gaussian by the central limit theorem. In the path-integral
formulation, commonly employed to study disordered systems, the Gaussian
approximation of $h_{k}$ is equivalent to a saddle-point approximation
to lowest order in the auxiliary field \citep[see e.g. ][eq. (3.5)]{Sompolinsky82_6860}\citep[eq. (3)]{Sompolinsky88_259}.
Even though one may expect the central limit theorem to be applicable
to the summed input $h_{k}$, it is easy to see that for networks
of several hundreds of units higher order cumulants still have an
effect. This is illustrated in \prettyref{fig:close_to_Gaussian_approximation},
showing the distribution of the input to a neuron receiving a sum
of binary, uncorrelated signals. The Gaussian approximation with the
same mean and variance as the exact distribution has a peak that is
slightly shifted to the left. Taking into account the third order
cumulant cures the displacement of the peak. The expansion in cumulants
in \prettyref{sec:Gaussian-approximation} shows how the higher order
cumulants can be taken into account in the calculation of the expectation
values of the gain function.

In the following we expand the distribution of $h_{k}$ up to third
order cumulants. To this end we use the relationship that the $n$-th
cumulant of a summed variable is the sum of $n$-th cumulants of its
constituents \eqref{eq:cumulant_sum}. We then use the property of
binary variables that $n_{k}^{K}=n_{k}$ for $K\geq1$, which relates
cumulants of order $K+L$ to cumulants of order $L+1$ in cases where
neuron $k$ appears $K$ times in the cumulant (see \prettyref{sub:Trivial-third-order}
for the detailed calculation). The first two cumulants of the summed
input $h_{k}=\sum_{l}J_{kl}n_{l}$ to a neuron are therefore given
as before by $\kappa_{1,k}=\mu_{k}$, $\kappa_{2,k}=\sigma_{k}^{2}$
\prettyref{eq:mu_sigma}. The third cumulant $\kappa_{3,k}$ (by defining
$x_{l}=J_{kl}n_{l}$ and $y=h_{k}$ and using \eqref{eq:cumulant_sum})
reads 
\begin{eqnarray}
\kappa_{3,k} & = & \sum_{ijr}J_{ki}J_{kj}J_{kr}\llangle n_{i}n_{j}n_{r}\rrangle,\label{eq:3rd_cumulant_input}
\end{eqnarray}
where we use the notation $\llangle\rrangle$ to denote cumulants.
For the mean activity \eqref{eq:stationary_statistics} we need to
evaluate the expectation value of a non-linear function applied to
$h_{k}$. We follow an analogous approach as in the previous section
(see \prettyref{sub:Functions-close-to-Gaussian} for details) and
apply \eqref{eq:<F(y)>_3rd} to $F_{k}(\n)=H(\sum_{j}J_{kj}n_{j}-\theta)$
yielding a perturbative treatment for the effect of the third order
cumulant 
\begin{eqnarray}
\langle F_{k}(\n)\rangle & = & e^{\frac{1}{6}\kappa_{3,k}\left(\frac{\partial}{\partial\mu_{k}}\right)^{3}}\,\langle H(h_{k}-\theta)\rangle_{h_{k}\sim\mathcal{N}(\mu_{k},\sigma_{k}^{2})}.\label{eq:mean_3rd_cum}
\end{eqnarray}
To obtain a correction of the covariances, we determine the contribution
of the third cumulant to the term of the form $\langle F_{k}(\n)\,n_{l}\rangle=\langle H(h_{k}(\n)-\theta)\,n_{l}\rangle$.
We apply the general result \prettyref{eq:<fx>} with $x_{l}=J_{kl}n_{l}$,
where we need to cancel a factor $J_{kl}$ in the final result due
to $n_{l}=x_{l}/J_{kl}$ to get 

\begin{eqnarray}
\langle F_{k}(\n)n_{l}\rangle & = & \sum_{q=0}^{3}\Delta\kappa_{q,kl}\frac{1}{q!}\,\left(\frac{\partial}{\partial\mu_{k}}\right)^{q}\,\langle F_{k}(\n)\rangle\label{eq:cond_gain_3rd}\\
\text{with}\nonumber \\
\Delta\kappa_{q,kl} & = & \sum_{i_{1}\cdots i_{q}}J_{ki_{1}}\cdots J_{ki_{q}}\llangle n_{i_{1}}\cdots n_{i_{q}}n_{l}\rrangle,\nonumber 
\end{eqnarray}
where $\langle F_{k}(\n)\rangle$ is given by \eqref{eq:mean_3rd_cum}.
The form of the terms $\Delta\kappa_{q,kl}\frac{1}{q!}\,\left(\frac{\partial}{\partial\mu_{k}}\right)^{q}=\Delta\kappa_{q,kl}\frac{\partial}{\partial\kappa_{q,k}}$
shows that, for $q\ge1$, they correspond to an infinitesimal displacement
of the $q$-th cumulant $\kappa_{q,k}$ by $\Delta\kappa_{q,kl}$.
These corrections come about by the presence of the variable $n_{l}$
in the expectation value, which can alternatively be understood as
a conditioned expectation value $\langle F_{k}(\n)n_{l}\rangle=m_{l}\langle F_{k}(\n)|n_{l}=1\rangle$,
where the condition $n_{l}=1$ changes the cumulants of $h_{k}$.
The first two terms in the sum \prettyref{eq:cond_gain_3rd} are identical
to the Gaussian approximation in \prettyref{eq:Gaussian_deriv_cumulant}.
The difference in the approximation schemes on the level of $\mathbf{n}$
and $h_{k}$, respectively, is apparent in the term for $q=2$, which
contains a correction to the second order cumulant due to the presence
of $n_{l}$. This term is neglected in the Gaussian truncation of
cumulants of $\mathbf{n}$ \prettyref{sec:Gaussian-approximation}.
The consistent approximation of $h_{k}$ up to third order cumulants
analogously requires the term for $q=3$. This correction in turn
depends on the fourth order cumulant $\llangle n_{i_{1}}\cdots n_{i_{q}}n_{l}\rrangle.$

We now use the properties of binary variables that allow us to express
a subset of higher cumulants by lower order ones due to the property
$\langle n_{i}^{K}\rangle=\langle n_{i}\rangle$ (for all integers
$K\ge1$). The $k$-th raw moment is given by the product of all combinations
of lower order cumulants \citep{VanKampen92}. For the third moment
this yields 
\begin{eqnarray}
\langle n_{l}n_{i}n_{j}\rangle & = & \llangle n_{l}n_{i}n_{j}\rrangle\nonumber \\
 & + & c_{li}m_{j}+c_{ij}m_{l}+c_{jl}m_{i}\label{eq:third_moment}\\
 & + & m_{l}m_{i}m_{j}.\nonumber 
\end{eqnarray}
If there are at least two identical indices, we can express the corresponding
third order cumulant by the two lower orders. Both cases $l=i\neq j$
and $l=i=j$ lead to the same expression \prettyref{eq:kappa_kkl_kkk}
\begin{eqnarray}
\llangle n_{l}n_{l}n_{j}\rrangle & = & c_{lj}(1-2m_{l}).\label{eq:kappa_llj}
\end{eqnarray}
In the latter case we get the third cumulant of a binary variable
$\llangle n_{l}n_{l}n_{l}\rrangle=m_{l}-3m_{l}^{2}+2m_{l}^{3}$ which
is uniquely determined by its mean. We can therefore take into account
the contribution of all third order cumulants (with at least two identical
indices) to the statistics of $h_{k}$. Stated differently, we calculate
the third order cumulant of $h_{k}$ using \prettyref{eq:3rd_cumulant_input}
and neglecting all cumulants of the binary variables where all indices
are different ($\llangle n_{i}n_{j}n_{l}\rrangle\simeq0$ for $i\neq j\neq l$).
A straight forward calculation (see \prettyref{sub:Trivial-third-order}
eq. \prettyref{eq:lambda_k_app} for details) yields

\begin{eqnarray}
\kappa_{3,k} & \simeq & \left[3(\mathbf{J}\circledast\mathbf{J})\,\diag(\{1-2m_{i}\})\,\mathbf{C}\mathbf{J}^{T}\right]_{kk}\nonumber \\
 & - & \left[2\left(\mathbf{J}\circledast\mathbf{J}\circledast\mathbf{J}\right)\,\diag\left(\{m_{i}-3m_{i}^{2}+2m_{i}^{3}\}\right)\right]_{k},\label{eq:lambda_k}
\end{eqnarray}
where we use the symbol $\circledast$ for the element-wise (Hadamard)
product of two matrices \citep{cichocki09}. For the third cumulant
appearing explicitly in \eqref{eq:cond_gain_3rd} we perform a similar
reduction that yields the matrix \prettyref{eq:eta_kl_app}

\begin{eqnarray}
\Delta\kappa_{2} & \simeq & \mathbf{J}\circledast\mathbf{J}\,\diag(\{1-2m_{i}\})\mathbf{C}\nonumber \\
 & + & 2\mathbf{J}\,\diag(\{1-2m_{i}\})\circledast(\mathbf{JC})\nonumber \\
 & - & 2(\mathbf{J}\circledast\mathbf{J})\,\diag\left(\{m_{i}-3m_{i}^{2}+2m_{i}^{3}\}\right).\label{eq:eta_kl}
\end{eqnarray}
The matrix $\Delta\kappa_{3}$ takes the form (see \prettyref{sub:Trivial-third-order}
for details)

\begin{eqnarray}
\Delta\kappa_{3} & \simeq & 3\mathbf{J}\circledast\left((\mathbf{J}\circledast\mathbf{J})\,\left\{ \llangle n_{i}n_{i}n_{j}n_{j}\rrangle_{ij}\right\} \right)\label{eq:Delta_kappa3}\\
 & + & 3(\mathbf{J}\circledast\mathbf{J})\circledast\left(\mathbf{J}\left\{ \llangle n_{i}n_{j}n_{j}n_{j}\rrangle_{ij}\right\} \right)\nonumber \\
 & + & (\mathbf{J}\circledast\mathbf{J}\circledast\mathbf{J})\left(\left\{ \llangle n_{i}n_{i}n_{i}n_{j}\rrangle_{ij}\right\} +\diag(\left\{ \llangle n_{i}n_{i}n_{i}n_{i}\rrangle_{i}\right\} )\right),\nonumber 
\end{eqnarray}
where the fourth order cumulants are given by \prettyref{eq:ni_ni_nk_nk}-\prettyref{eq:ni_ni_ni_ni}.

To evaluate the expression \eqref{eq:mean_3rd_cum} for the mean activity
and \eqref{eq:cond_gain_3rd} for the covariance, we need the $n$-th
derivative of the complementary error function. We use that $\frac{d}{dx}\,\frac{1}{2}\erfc(-x)=\frac{1}{\sqrt{\pi}}\,e^{-x^{2}}$
is a Gaussian and further that the $n$-th derivative of a Gaussian
$\left(\frac{d}{dx}\right)^{n}e^{-x^{2}}=(-1)^{n}H_{n}(x)\,e^{-x^{2}}$
can be expressed in terms of the $n$-th Hermite polynomial $H_{n}$
\citep[ 18.3 "physicists notation"]{DLMF12}. Each differentiation
with respect to $\mu_{k}$ yields an additional factor $\left(\sqrt{2}\sigma_{k}\right)^{-1}$.
Hence we get for $n\ge1$
\begin{eqnarray*}
L_{n}(\mu_{k},\sigma_{k}) & := & \left(\frac{d}{d\mu_{k}}\right)^{n}\frac{1}{2}\erfc\left(\frac{\theta-\mu_{k}}{\sqrt{2}\sigma_{k}}\right)\\
 & = & \frac{1}{\sqrt{\pi}}\,\left(\sqrt{2}\sigma_{k}\right)^{-n}\left.H_{n-1}\left(x\right)\,e^{-x^{2}}\right|_{x=\frac{\theta-\mu_{k}}{\sqrt{2}\sigma_{k}}}.
\end{eqnarray*}
Using this definition and the expansion $e^{\frac{1}{6}\kappa_{3,k}\,\left(\frac{\partial}{\partial\mu_{k}}\right)^{3}}\simeq1+\frac{1}{6}\kappa_{3,k}\,\left(\frac{\partial}{\partial\mu_{k}}\right)^{3}$,
valid for $\kappa_{3,k}$ small compared to $\sigma_{k}^{3}$, we
get the mean activity
\begin{eqnarray}
m_{k}=\langle F_{k}(\n)\rangle & \simeq & \left(1+\frac{1}{6}\kappa_{3,k}\,\left(\frac{\partial}{\partial\mu_{k}}\right)^{3}\right)\frac{1}{2}\erfc\left(\frac{\theta-\mu_{k}}{\sqrt{2}\sigma_{k}}\right)\nonumber \\
 & = & L_{0}(\mu_{k},\sigma_{k})+\frac{1}{6}\kappa_{3,k}\,L_{3}(\mu_{k},\sigma_{k}).\label{eq:mean_activity_3rd}
\end{eqnarray}
For the term appearing in the covariance we get

\begin{eqnarray}
\langle F_{k}(\n)\,n_{l}\rangle & \simeq & m_{l}m_{k}\nonumber \\
 & + & \sum_{j=1}^{N}\left(L_{1}(\mu_{k},\sigma_{k})+\frac{1}{6}\kappa_{3,k}\,L_{4}(\mu_{k},\sigma_{k})\right)\,J_{kj}c_{jl}\nonumber \\
 & + & \frac{1}{2}\left(L_{2}(\mu_{k},\sigma_{k})+\frac{1}{6}\kappa_{3,k}\,L_{5}(\mu_{k},\sigma_{k})\right)\,\Delta\kappa_{2,kl}\nonumber \\
 & + & \frac{1}{6}\left(L_{3}(\mu_{k},\sigma_{k})+\frac{1}{6}\kappa_{3,k}\,L_{6}(\mu_{k},\sigma_{k})\right)\,\Delta\kappa_{3,kl}\nonumber \\
 & + & o(\kappa_{3,k}^{2}).\label{eq:cond_F_3rd_final}
\end{eqnarray}
The latter expression shows in particular that the correction to the
mean activity reappears (as the factor $m_{k}$) in the first line,
indicating that the contribution of the third cumulant to lowest order
($q=0$ in \eqref{eq:cond_gain_3rd}) drops out of the covariance,
as the term $m_{l}m_{k}$ is subtracted in \eqref{eq:stationary_statistics}.
In a weakly correlated state of the network, the remaining terms are
smaller than $m_{l}m_{k}$ as they give rise to the pairwise covariance
\eqref{eq:stationary_statistics}. This explains why the Gaussian
approximation is already fairly accurate.

\begin{figure}
\begin{centering}
\includegraphics{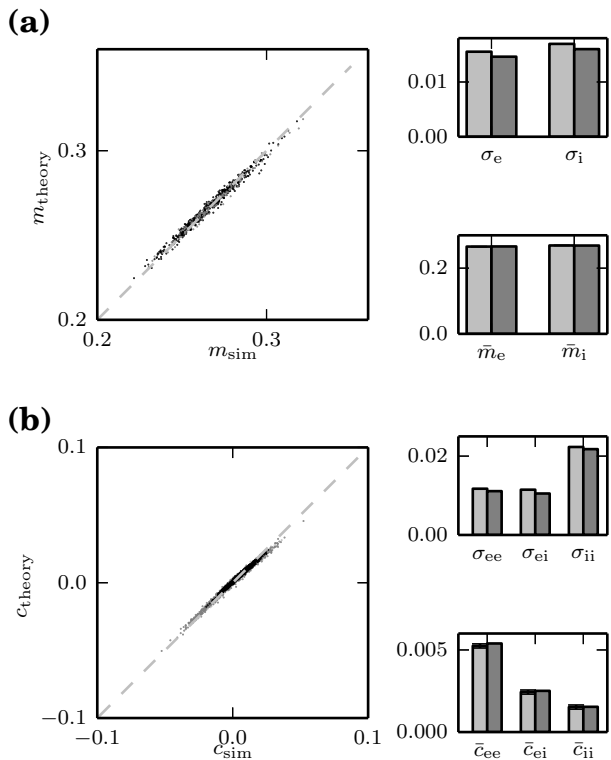}
\par\end{centering}

\caption{\textbf{Approximation for covariance matrix in random network of binary
neurons including third order cumulants.} \textbf{(a) }Theoretical
prediction of mean activities \eqref{eq:mean_activity_3rd} versus
simulation results (black: excitatory, gray: inhibitory). Lower inset:
mean activity averaged over excitatory and inhibitory neurons (light
gray: simulation, darker gray: theory). Black error bars (not visible,
below line width) show standard error of the mean obtained from $20$
simulations. Upper inset: standard deviation of distribution of mean
activities of excitatory ($\sigma_{e}$) and inhibitory ($\sigma_{i}$)
neurons. \textbf{(b)} Theoretical prediction \prettyref{eq:cond_F_3rd_final}
versus simulated covariance (black: excitation-excitation, gray: inhibition-inhibition,
light gray: excitation-inhibition). Lower inset: mean covariances
averaged over excitatory-excitatory, excitatory-inhibitory and over
inhibitory-inhibitory neuron pairs (light gray: simulation, darker
gray: theory). Error bars show standard error of the mean obtained
from $20$ simulations. Upper inset: standard deviations of the distributions
of covariances. Network parameters and display as in \prettyref{fig:covariance_lin}.
Theoretical predictions obtained by damped fixed-point iteration.\label{fig:covariance_nonlin}}
\end{figure}

Simultaneously solving \eqref{eq:mean_activity_3rd} and \eqref{eq:cond_F_3rd_final}
by a damped fixed point iteration, analogous to \eqref{eq:damped_fixed_point},
yields an approximation of the mean activities and covariances shown
in \prettyref{fig:covariance_nonlin}. The deviation of the mean activities
(\prettyref{fig:covariance_nonlin}a) from simulation results is reduced
compared to the Gaussian approximation below significance level. The
width of the distributions is only slightly underestimated compared
to simulations, as exhibited by the scatter plot aligned to the diagonal
and the bar graph. The pairwise averaged cross-covariances (\prettyref{fig:covariance_nonlin}b)
are within the error of the simulated results, in contrast to the
Gaussian approximation (cf. \prettyref{fig:covariance_lin}b). A small
contribution to the remaining difference in variance stems from the
finite simulation time, naturally leading to a wider distribution
in \prettyref{fig:covariance_nonlin}b compared to the theoretical
prediction. The bigger contribution presumably comes from the non-trivial
third order cumulants $\llangle n_{i}n_{j}n_{k}\rrangle$, where all
indices are unequal. Neglecting the non-trivial covariances shows
that the variability of $\sigma_{k}$ from neuron to neuron is reduced,
which in turn reduces the width of the distribution of the mean activities.
For networks with fixed in-degree this even leads to a uniform mean
activity across neurons, which, however, still matches the averaged
mean activities from simulations indicating that averaged quantities
are insensitive to variability in non-trivial higher-order cumulants
(see \prettyref{fig:Lyapunov} in \prettyref{sec:Solution-of-the-modufied-Lyapunov}).
Analogously we expect that the neglect of non-trivial third order
cumulants underlies the reduced width of the covariances.

\section{Scale invariance of covariances\label{sec:scale_invariance}}

The equation for cross-covariances in \eqref{eq:lin_sol} yields an
additional insight: Assuming that mean activities and correlations
were unchanged, a scaling of all incoming synapses to a neuron $k$
by some factor $\alpha>0$ amounts to a scaling of the strength of
synaptic fluctuations in the same manner ($\sigma_{k}\propto\alpha$).
The fixation of mean activities can be achieved for different choices
of incoming synaptic amplitudes by adapting the threshold such that
$\left(\mu_{k}(\{J_{kl}\})-\theta_{k}\right)/\sigma_{k}$ remains
invariant (cf. \eqref{eq:lin_sol}). A uniform rescaling of all synapses
by a factor $\alpha>0$ therefore requires a change of the threshold
by the same factor. The susceptibility \eqref{eq:susceptibility-1}
$S_{k}$ then scales as $\sigma_{k}^{-1}\propto\alpha^{-1}$. Hence
the term $S_{k}J_{kl}$ appearing in the equation for covariances
in \eqref{eq:lin_sol} is invariant under this scaling. This implies
that covariances are invariant with respect to the absolute value
of synaptic amplitudes: The self-generated network noise causes a
divisive normalization on the level of the synaptic input to each
neuron. This also implies that scaling the synapses with $J\propto N^{-1}$,
$J\propto N^{-\frac{1}{2}}$, or $J\propto1$ as the network size
tends to infinity all yield the same covariances, given the thresholds
are adapted so that the mean activity is preserved. On the level of
population-averaged covariances this has been remarked earlier \citep{Grytskyy13_258}.
The independence of the network dynamics on the absolute value of
the synaptic weights even holds exactly, as noted earlier \citep{Albada15}.
The reason is the absence of a length scale of a hard threshold; the
only relevant length scale is the amplitude $\sigma_{k}$ of the synaptic
noise itself. Considering a single neuron $k$, the condition for
the neuron to be activated is $h_{k}=\sum_{l}J_{kl}n_{l}>\theta_{k}$.
Rescaling all incoming synapses as well as the threshold by the same
factor $\alpha>0$ multiplies both sides of this inequality; the neuron
switches at the same configuration $\n$ of incoming spins as in the
original case. This consideration is completely in line with the work
of \citep{Vreeswijk98}. The latter work found that, when increasing
the number of neurons $N$, a scaling of $1/\sqrt{N}$ is required
to obtain a robust asynchronous state, if the system possesses variability
of the thresholds on the scale defined as unity. The width of the
distribution of the thresholds induces a second length scale into
the system. Invariant behavior can therefore only arise, if the synaptic
noise $\sigma_{k}$ and the standard deviation of the thresholds have
a constant ratio. Scaling synapses as $J\propto1/N$ in this setting
leads to vanishing temporal fluctuations of $h_{k}$ and hence spin
glass freezing, whereas for $J\propto1$ the asynchronous state persists.
Fixing $J$, while increasing the number of neurons $N$, results
in more inputs per neuron and thus increased temporal fluctuations,
which wash out the effect of distributed thresholds.

\section{Mapping of fluctuations to Ornstein-Uhlenbeck processes\label{sec:Solution-of-the-modufied-Lyapunov}}

We here show an alternative interpretation of the Gaussian approximation
in \prettyref{sec:Gaussian-approximation}. Despite the fact that
the covariances obey different equations on the diagonal $k=l$ and
the off-diagonal $k\neq l$ \eqref{eq:lin_sol}, we can rewrite the
equations as a single matrix equation ($\mathbf{W}=\mathbf{S}\mathbf{J}$,
$\mathbf{S}=\diag(S_{1},\ldots,S_{N})$) 
\begin{eqnarray}
2\mathbf{C}-\mathbf{D} & = & \mathbf{WC}+(\mathbf{WC})^{T}\label{eq:cov_linear}\\
0 & = & (\mathbf{W}-\mathbf{1})\mathbf{C}+((\mathbf{W}-\mathbf{1})\mathbf{C})^{T}+\mathbf{D},\nonumber 
\end{eqnarray}
where $\mathbf{D}$ is a diagonal matrix with elements constrained
by the condition that the diagonal entries of the covariance matrix
fulfill \eqref{eq:stationary_statistics}
\begin{eqnarray}
c_{kk} & = & m_{k}(1-m_{k}).\label{eq:autocov_binary}
\end{eqnarray}
We ensure this latter condition in the following way. We will first
solve \eqref{eq:cov_linear} by multiplication with the left eigenvectors
$\mathbf{v}_{\alpha}$ of the connectivity, i.e. $\mathbf{v}_{\alpha}^{T}(\mathbf{W}-\mathbf{1})=\lambda_{\alpha}\mathbf{v}_{\alpha}^{T}$
\citep[chapter 6.5]{Risken96}. The corresponding right eigenvectors
are $\mathbf{u}_{\alpha}$, which fulfill the bi-orthogonality $\mathbf{v}_{\alpha}^{T}\mathbf{u}_{\beta}=\delta_{\alpha\beta}$
and completeness $\sum_{\alpha}\mathbf{u}_{\alpha}\mathbf{v}_{\alpha}^{T}=1$
relation. With the notation $C^{\alpha\beta}:=\mathbf{v}_{\alpha}^{T}\mathbf{C}\mathbf{v}_{\beta}$
(analogously for $\mathbf{D}$) we have
\begin{eqnarray}
C^{\alpha\beta} & = & -\frac{D^{\alpha\beta}}{\lambda_{\alpha}+\lambda_{\beta}}\nonumber \\
\mathbf{C} & = & \sum_{\alpha\beta}\mathbf{u}_{\alpha}\mathbf{u}_{\beta}^{T}\,C^{\alpha\beta},\label{eq:C_backtrafo}
\end{eqnarray}
where the latter expression follows from the completeness relation.

The expression reveals the connection between the eigenvalues $\lambda_{\alpha}$
of the linearized coupling matrix and the fluctuations in the system.
In the case of a single eigenvalue close to instability, i.e. $\Re(\lambda_{\alpha})\simeq0$,
the fluctuations in the corresponding eigendirection constitute the
dominant contribution to the covariance matrix $\mathbf{C}\simeq-\mathbf{u}_{\alpha}\mathbf{u}_{\alpha}^{T}\,\frac{\mathbf{v}_{\alpha}^{T}\mathbf{D}\mathbf{v}_{\alpha}}{2\lambda_{\alpha}}$.
This scenario could experimentally be detected in a principle component
analysis, with the dominant principle component pointing in the direction
of $\mathbf{u}_{\alpha}$.

Evaluating expression \eqref{eq:C_backtrafo} on the diagonal and
using $D^{\alpha\beta}=v_{\alpha}^{T}\mathbf{D}v_{\beta}=\sum_{j}v_{\alpha,j}v_{\beta,j}D_{j}$
we get
\begin{align}
m_{k}(1-m_{k}) & \stackrel{!}{=}c_{kk}=\sum_{j}\underbrace{-\left(\sum_{\alpha\beta}\,\frac{u_{\alpha,k}v_{\alpha,j}\cdot u_{\beta,k}v_{\beta,j}}{\lambda_{\alpha}+\lambda_{\beta}}\right)}_{\equiv B_{kj}}D_{j}\label{eq:noise_Lyapunov}\\
\mathbf{B} & =-\sum_{\alpha\beta}\frac{(\mathbf{u}_{\alpha}\mathbf{v}_{\alpha}^{T})\circledast(\mathbf{u}_{\beta}\mathbf{v}_{\beta}^{T})}{\lambda_{\alpha}+\lambda_{\beta}},\nonumber 
\end{align}
where the symbol $\circledast$ is to be understood as the element-wise
multiplication (Hadamard product) of the two matrices in the numerator.
The penultimate line is an ordinary matrix equation relating the ($N$
dimensional vector) $D_{k}$ to the ($N$ dimensional vector) $c_{kk}$.
To determine $\mathbf{D}$ as $\mathbf{D}=\mathbf{B}^{-1}\diag(\{c_{kk}\})$
we hence need to invert the matrix $\mathbf{B}$. The covariance matrix
is then obtained by \eqref{eq:C_backtrafo}.

The result \eqref{eq:cov_linear} can be further interpreted as a
mapping of the fluctuations from the binary dynamics to an effective
system of Ornstein-Uhlenbeck processes. Given the set of coupled Ornstein-Uhlenbeck
processes
\begin{eqnarray}
\tau\frac{\partial x_{k}(t)}{\partial t} & = & -x_{k}(t)+\sum_{j}w_{kj}x_{j}(t)+\xi_{k}(t)\label{eq:OUP_fluct}
\end{eqnarray}
with the Gaussian white noise $\langle\mathbf{\xi}(t)\mathbf{\xi}^{T}(s)\rangle=\tau\mathbf{D}\,\delta(t-s)$,
the stationary equal-time covariance matrix fulfills the same continuous
Lyapunov equation \citep[chapter 6.5]{Risken96} as the binary network
\prettyref{eq:cov_linear}. By the analogy of the continuous Lyapunov
equations we see that the elements of the diagonal matrix $\mathbf{D}$
can be interpreted as the noise intensity injected into each neuron.
The modified Lyapunov method \prettyref{eq:noise_Lyapunov} determines
this intensity such that the variance of each continuous variable
$x_{k}$ agrees to that of the corresponding binary variabe $n_{k}$
given by \eqref{eq:autocov_binary} that, in turn, is fixed by the
mean activity \eqref{eq:lin_sol}.

\begin{figure}
\begin{centering}
\includegraphics{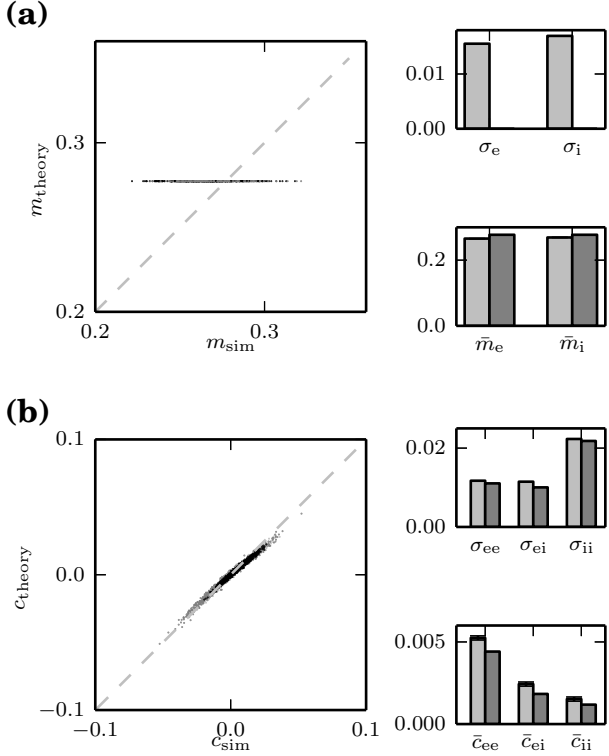}
\par\end{centering}

\caption{\textbf{Modified Lyapunov method.} Solution of the individual pairwise
covariances by the projection method applied to the continuous Lyapunov
equation in comparison to simulation results. \textbf{(a) }Mean activities
predicted by the approximation that neglects cross-correlations (black:
excitatory, gray: inhibitory). Lower inset: mean activity averaged
over excitatory and inhibitory neurons (light gray: simulation, darker
gray: theory). Black error bars show the standard error of the mean
obtained from $20$ simulations (barely visible, approximately at
line width). Upper inset: width of distribution of mean activities.\textbf{
(b) }Theoretical distribution of covariances \eqref{eq:C_backtrafo}
with $\mathbf{D}$ chosen according to \eqref{eq:noise_Lyapunov}
(same approximation as in (a) for $\mathbf{m}$ and $\mathbf{S}$,
i.e. neglect of cross-correlations in the input to each neuron). Lower
inset: mean covariances averaged over excitatory-excitatory, excitatory-inhibitory
and over inhibitory-inhibitory neuron pairs (light gray: simulation,
darker gray: theory). Error bars show the standard error of the mean
obtained from $20$ simulations. Upper inset: width of the distributions
of covariances. Other parameters and display as in \prettyref{fig:covariance_lin}.\label{fig:Lyapunov}}
\end{figure}

It is important to note that $\mathbf{W}$ and $\mathbf{D}$ in \eqref{eq:cov_linear}
and \eqref{eq:OUP_fluct} themselves depend on the covariances $\mathbf{C}$
via the susceptibility $\mathbf{S}$. This leads to \eqref{eq:C_backtrafo}
remaining an implicit equation for $\mathbf{C}$ that needs to be
solved iteratively. However, by neglecting the contribution of cross-covariances
in \prettyref{eq:mu_sigma}, $\mathbf{W}$ and $\mathbf{D}$ become
independent of $\mathbf{C}$, rendering \eqref{eq:cov_linear} a linear
equation for the covariances and the above projection method an efficient
algorithm to compute $\mathbf{C}$. This linear approximation consequently
fails to predict the distribution of mean activities for networks
with fixed indegree, as shown in \prettyref{fig:Lyapunov}a. Nevertheless,
the width of the distribution of the covariances shown in \prettyref{fig:Lyapunov}b
is only slightly underestimated, showing that the distribution of
mean activities contributes only marginally to the distribution of
covariances.

The viability of the linear approximation for the covariances shows
that fluctuations in the binary model are practically equivalent to
those in linear Ornstein-Uhlenbeck processes. This equivalence has
been reported earlier for effective equations of the population-averaged
pairwise correlations \citep{Grytskyy13_131}. The latter work used
the approximate result $c_{kk}\simeq\frac{D_{k}}{2}$, which ignores
the effect of the covariances onto the auto-covariances in \eqref{eq:cov_linear},
i.e. assuming that $(\mathbf{WC})_{kk}$ is smaller than $c_{kk}$
itself and hence neglecting the right hand side of \eqref{eq:cov_linear}
when determining the diagonal elements $c_{kk}$. In the absence of
self-connections this amounts to neglecting the off-diagonal cross-covariances
in $\mathbf{C}$. For weakly correlated network states this is a good
approximation. However, in the present network setting it slightly
overestimates the population-averages of the covariances as well as
the width of their distribution (data not shown).

\section{Response of the network to external stimuli\label{sec:Response-of-network}}

To study how the recurrent network processes an externally applied
signal, we take into account an additional external input to each
neuron $k$, denoted as $h_{k,\mathrm{ext.}}(t)$, such that $h_{k}(t)=\sum_{j}J_{kj}n_{j}(t)+h_{k,\mathrm{ext.}}(t)$.
We start from the differential equation \eqref{eq:1st_2nd_moment}
for the mean activities 
\begin{align*}
\tau\frac{\partial}{\partial t}\langle n_{k}(t)\rangle & =-\langle n_{k}(t)\rangle+\langle F_{k}(\n(t),h_{k,\mathrm{ext.}}(t))\rangle,
\end{align*}
with

\begin{eqnarray*}
F_{k}(\n(t),h_{k,\mathrm{ext.}}(t)) & = & H\left(\sum_{j}J_{kj}n_{j}(t)+h_{k,\mathrm{ext.}}(t)-\theta\right).
\end{eqnarray*}
As in the case without external input, the expectation value $\langle F_{k}(\n(t),h_{k,\mathrm{ext.}}(t))\rangle$
can be treated in the Gaussian approximation. Subtracting the stationary
activity state $\delta n_{k}(t)=n_{k}(t)-\langle n_{k}\rangle$ we
get after linearization

\begin{eqnarray*}
\tau\frac{\partial}{\partial t}\langle\delta n_{k}(t)\rangle & = & -\langle\delta n_{k}(t)\rangle+S_{k}\left(\sum_{j}J_{kj}\langle\delta n_{j}(t)\rangle+h_{k,\mathrm{ext.}}(t)\right)\\
 & = & \sum_{j}(w_{kj}-\delta_{kj})\langle\delta n_{j}(t)\rangle+\underbrace{S_{k}\,h_{k,\mathrm{ext.}}(t)}_{=:y_{k}(t)},
\end{eqnarray*}
where $w_{kj}=S_{k}J_{kj}$ is the effective connectivity as defined
in \prettyref{sec:Solution-of-the-modufied-Lyapunov} and $S_{k}$
is the susceptibility \eqref{eq:susceptibility-1}. Hence the equation
of motion for the perturbation in matrix notation reads

\begin{eqnarray}
\tau\frac{\partial}{\partial t}\langle\delta\mathbf{n}(t)\rangle & = & (\mathbf{W}-\mathbf{1})\langle\delta\mathbf{n}(t)\rangle+\mathbf{y}(t).\label{eq:equ_mo_pert}
\end{eqnarray}
In order to excite the system into the direction of one eigenmode
of the effective connectivity $\mathbf{W}$, we choose the following
stimulus vector 

\begin{eqnarray}
\mathbf{h}_{\mathrm{ext.}}^{\alpha}(t) & := & a\,\tilde{\mathbf{u}}_{\alpha}\,f(t),\:\mbox{with}\:\tilde{\mathbf{u}}_{\alpha}=\frac{\mathbf{S}^{-1}\Re(\mathbf{u}_{\alpha})}{\|\mathbf{S}^{-1}\Re(\mathbf{u}_{\alpha})\|},\label{eq:normalized_u-1-1-1}
\end{eqnarray}
where $\mathbf{S}:=\diag(\{S_{k}\})$ is the diagonal matrix containing
the susceptibilities and $\mathbf{u}_{\alpha}$ the right-sided eigenvector
of $\mathbf{W}-\mathbf{1}$ as defined in \prettyref{sec:Solution-of-the-modufied-Lyapunov}.
The strength of the stimulus is regulated by the parameter $a$ and
its temporal profile determined by $f(t)$. The parameter $a$ allows
us to control the amplitude of the stimulation in comparison to the
strength of the synaptic noise received by each unit. For the linear
approximation to hold, this amplitude must be chosen such that the
input to each unit is in the linear part of the expectation value
of the gain function and can hence be approximated by the slope \eqref{eq:susceptibility-1}.
Stimulating into the direction of the real part of the eigenvectors
ensures that a complex mode is excited in combination with its complex
conjugate, which is necessary since the activity in the network is
real-valued. Here, $\tilde{\mathbf{u}}_{\alpha}$ is constructed such
that it is normalized and compensates for the multiplication of the
external input with the diagonal susceptibility matrix. We measure
the deflection of the activity into the direction of the eigenmode
by defining $\langle\delta n^{\alpha}(t)\rangle:=\tilde{\mathbf{v}}_{\alpha}^{T}\langle\delta\mathbf{n}(t)\rangle$
as the projection of the activity vector onto the $\alpha$-th eigenmode
of the connectivity matrix $\mathbf{W}$ ($y^{\alpha}$ defined analogously),
where we choose $\tilde{\mathbf{v}}_{\alpha}^{T}$ such that $\tilde{\mathbf{v}}_{\alpha}^{T}\mathbf{S}\tilde{\mathbf{u}}_{\beta}=\delta_{\alpha\beta}$,
i.e. 

\begin{eqnarray*}
\mathbf{\tilde{\v}}_{\alpha} & :=k\|\mathbf{S}^{-1}\Re(\mathbf{u}_{\alpha})\|\Re(\v_{\alpha}),\:\mbox{with}\:k= & \begin{cases}
1 & \quad\text{if }\Im(\lambda_{\alpha})=0\\
2 & \quad\text{if }\Im(\lambda_{\alpha})\neq0
\end{cases}
\end{eqnarray*}
with $\mathbf{v}_{\alpha}$ being the left eigenvectors of $\mathbf{W}-\mathbf{1}$
as defined in \prettyref{sec:Solution-of-the-modufied-Lyapunov}.

\begin{figure}
\centering{}\includegraphics{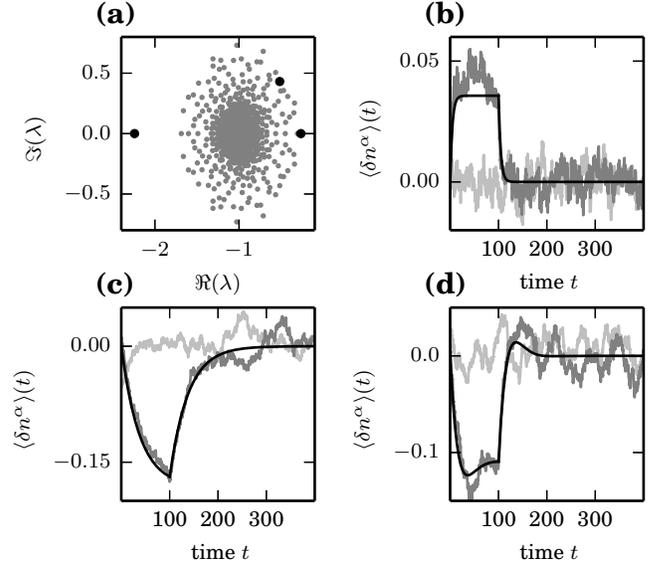}\caption{\textbf{Linear responses to DC-stimuli.} Responses (normalized by
the number of units $N$) to a DC-stimulus (duration $T=100\protect\ms$)
in the direction of one eigenmode (dark gray), the background activity
(light gray) measured in simulation and the analytical prediction
of the response curve (\prettyref{eq:response_final}, black) obtained
from linear response theory. \textbf{(a) }Eigenvalues $\lambda_{\alpha}$
of the matrix $\mathbf{W}-\mathbf{1}$. The eigenvalues associated
with the modes that are stimulated are depicted by black dots. \textbf{(b)
}Stimulation in the direction of the fastest decaying eigenmode $\lambda=-2.24$,
with a stimulus size $a=50$. \textbf{(c)} Stimulation in the direction
of the slowest decaying eigenmode $\lambda=-0.26$, with a stimulus
size $a=-30$. \textbf{(d)} Stimulation in the direction of an eigenmode
decaying slowly and oscillatory $\lambda=-0.52+0.43i$, with a stimulus
size $a=-60$. Responses averaged over $500$ repetitions. Other parameters
as in \prettyref{fig:covariance_lin}.\label{fig:Linear-responses}}
\end{figure}

The time evolution of the perturbation $\langle\delta n^{\alpha}(t)\rangle$
is obtained by solving \eqref{eq:equ_mo_pert} projected onto $\v_{\alpha}$
and $\v_{\alpha}^{*}$, and subsequently adding the results. Hence,
for any stimulus $f(t)$ (inserted into \eqref{eq:equ_mo_pert} as
$\mathbf{y}(t)=\mathbf{S}\mathbf{h}_{\mathrm{ext.}}^{\alpha}(t)$)
the perturbation obeys the convolution equation

\begin{eqnarray}
\langle\delta n^{\alpha}\rangle(t) & = & a\,\frac{1}{\tau}\,\frac{1}{2}\,\left[\left(e^{\lambda_{\alpha}\frac{\circ}{\tau}}+e^{\lambda_{\alpha}^{\ast}\frac{\circ}{\tau}}\right)\ast f\right](t).\label{eq:response_complex-1}
\end{eqnarray}
We here consider an incoming DC-input starting at time $t_{0}=0$
and stopping at $t=T$, i.e. we choose $f(t)=H(t)-H(t-T)$. The time
course of the perturbation is then given by

\begin{equation}
\langle\delta n^{\alpha}\rangle(t)=a\begin{cases}
\Re\Big(\lambda_{\alpha}^{-1}(1-e^{\lambda_{\alpha}\frac{t}{\tau}})\Big) & \quad\text{if }t\leq T\\
\Re\Big(\lambda_{\alpha}^{-1}(e^{\lambda_{\alpha}\frac{t-T}{\tau}}-e^{\lambda_{\alpha}\frac{t}{\tau}})\Big) & \quad\text{if }t>T
\end{cases}.\label{eq:response_final}
\end{equation}
\prettyref{fig:Linear-responses} shows the response of the network
activity projected onto one eigenmode to the DC-stimulus in the direction
of the same eigenmode. Choosing three stimulus directions associated
with three representative eigenvalues from the full eigenvalue cloud
of the network (\prettyref{fig:Linear-responses}a), we demonstrate
that the time course of fast (large negative eigenvalue), slowly (eigenvalue
close to zero), as well as oscillatory (complex eigenvalue) decaying
modes is captured by the linear approximation. This shows how the
response of the network depends on the spatial structure of the stimulation.
To lowest order, the transformation performed by the network is hence
a spatio-temporal filtering of the input, where the responses with
slowest decay correspond to the excitation of eigenmodes closest to
instability.

\section{Reconstruction of connectivity\label{sec:Slope-of-correlation}}

\begin{figure}
\begin{centering}
\includegraphics{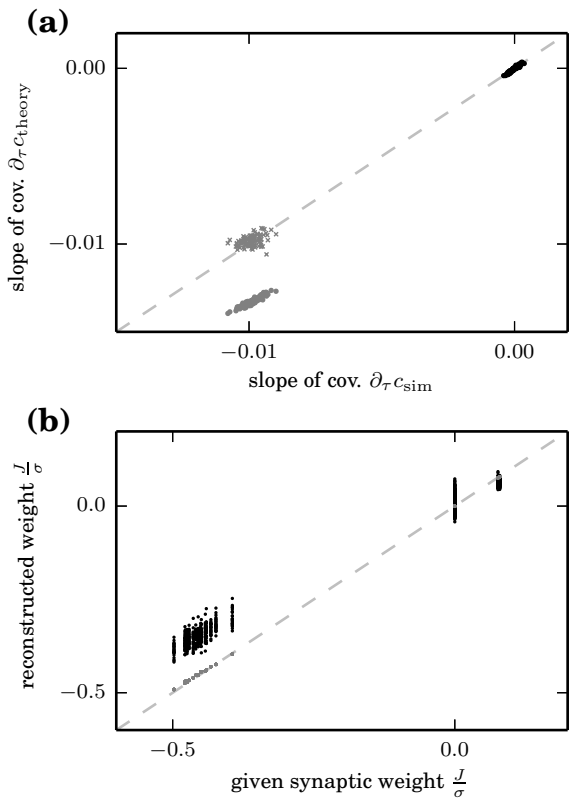}
\par\end{centering}

\caption{\textbf{Reconstruction of connectivity from the matrix of covariances
and slopes of the covariance functions at zero time lag.} \textbf{(a)}
Slope of auto- (gray) and cross-covariance functions (black) at zero
time lag from the Gaussian approximation versus slope taken from simulation
results. Dots show the prediction of the linear theory \prettyref{eq:slope_OUP},
crosses the approximation $-\frac{1}{2}m_{k}(1-m_{k})$ for the slope
of auto-covariances. \textbf{(b)} Reconstructed synaptic amplitude
$w_{kl}$ in a network of Ornstein-Uhlenbeck processes \prettyref{eq:reconstruction_OUP}
(gray) and reconstructed weights $J_{kl}/\sigma_{k}$ \prettyref{eq:reconstructed_weight}
(black) in a binary network versus original couplings used in simulation.
Covariance and slope averaged over $20$ repetitions, each simulated
for $T=2,000,000\protect\ms$. Other network parameters as in \prettyref{fig:covariance_lin}.\label{fig:Reconstruction-of-connectivity}}
\end{figure}

In the current section we investigate the inverse problem, i.e. in
how far the connectivity $J_{kl}$ of the network can be inferred
from the observed activity. The time-lagged cross-covariance function
in a network of Ornstein-Uhlenbeck processes (for $t>s$) fulfills
the differential equation \citep[ chapter 6.5]{Risken96}

\begin{eqnarray}
q_{kl}:=\tau\frac{\partial}{\partial t}c_{kl}(t,s) & +c_{kl}(t,s)\stackrel{\text{\ensuremath{{\scriptstyle \mbox{\ensuremath{t>s}}}}}}{=} & \sum_{j}w_{kj}c_{jl}(t,s).\label{eq:slope_OUP}
\end{eqnarray}
This means we can uniquely reconstruct the connectivity $w_{kl}$
from the knowledge of the two matrices, the covariance and its slope
at time lag zero as
\begin{eqnarray}
\mathbf{W} & = & \mathbf{Q}\mathbf{C}^{-1}\nonumber \\
 & = & \mathbf{1}+\tau\frac{\partial}{\partial t}\left.\mathbf{C}(t,s)\right|_{t=s}\mathbf{C}^{-1}.\label{eq:reconstruction_OUP}
\end{eqnarray}
For an Ornstein-Uhlenbeck process, obeying \prettyref{eq:cov_linear}
and \prettyref{eq:slope_OUP}, the reconstruction of the connectivity
$w_{kl}$ is exact, as shown in \prettyref{fig:Reconstruction-of-connectivity}b.

\begin{figure}
\begin{centering}
\includegraphics{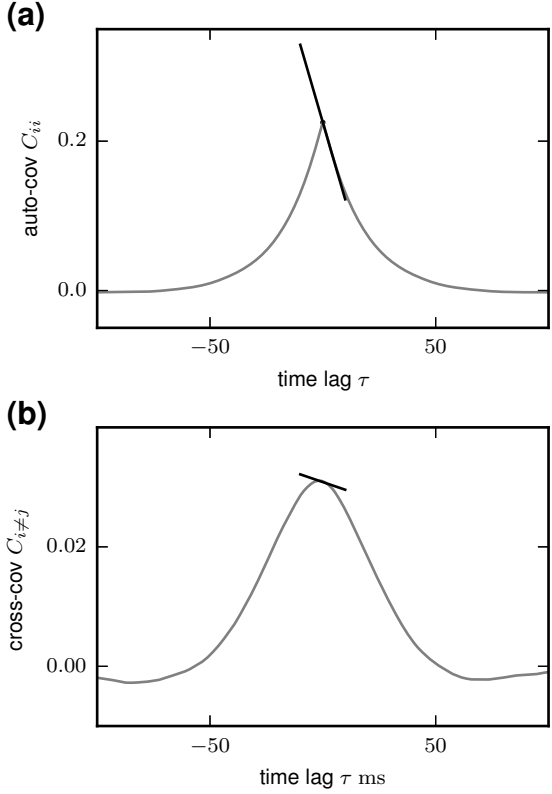}
\par\end{centering}

\caption{\textbf{Auto- and cross covariance functions.} Example of an auto-
(\textbf{a})\textbf{ }and cross-covariance function (\textbf{b})\textbf{
}in the network described in \prettyref{fig:covariance_lin}. The
respective slopes of the functions at time lag $0+$ are indicated
by the black tangent lines.\label{fig:auto_cross_cov_fctn}}
\end{figure}

We will now demonstrate that a similar relationship approximately
also holds in binary networks. For $t>s$, the time-lagged cross-covariance
function $c_{kl}(t,s)=\langle n_{k}(t)n_{l}(s)\rangle-\langle n_{k}(t)\rangle\langle n_{l}(t)\rangle$
for the binary network fulfills \prettyref{eq:time_lagged_correlation}
\begin{eqnarray}
q_{kl}(t,s) & := & \tau\frac{\partial}{\partial t}c_{kl}(t,s)+c_{kl}(t,s)\nonumber \\
 & \stackrel{\text{\ensuremath{{\scriptstyle \mbox{\ensuremath{t>s}}}}}}{=} & \langle F_{k}(\n(t))\,n_{l}(s)\rangle-\langle n_{k}(t)\rangle\langle n_{l}(s)\rangle.\label{eq:def_D}
\end{eqnarray}
An example of an auto- and cross-covariance function together with
the slope at zero time lag is shown in \prettyref{fig:auto_cross_cov_fctn}.

In the limit of vanishing time lag the form of $q_{kl}=\lim_{t\to s}q_{kl}(t,s)=\langle F_{k}(\n(t))\,n_{l}(t)\rangle-\langle n_{k}(t)\rangle\langle n_{l}(t)\rangle$
shows that $q_{kl}$ measures the direct influence of fluctuations
of neuron $l$ on the gain function of neuron $k$. In the Gaussian
approximation (see \prettyref{sec:Gaussian-approximation}) we get
for $k\neq l$
\begin{eqnarray}
q_{kl} & = & \sum_{j}S_{k}J_{kj}c_{jl}.\label{eq:q_kl}
\end{eqnarray}
For $k=l$, however, we have to evaluate $\langle H(h_{k}(\n(t)-\theta)\,n_{k}(t)\rangle$
on the right hand side. Since the term $n_{k}$ appears in the expectation
value, the statistics of $h_{k}$ is effectively conditioned on the
state $n_{k}=1$. Since the transitions of neuron $k$ directly depend
on the value of $h_{k}$, this conditioning violates the close-to
Gaussian approximation of $h_{k}|n_{k}=1$. An approximate treatment
on the diagonal is possible under the assumption that the auto-covariance
function of $h_{k}$ is dominated by contributions of the auto-covariances
of the binary variables $n$, yielding a non-linear differential equation
for the temporal shape of the auto-covariance function \citep[ eq. (5.17)]{Vreeswijk98}.
Approximating the temporal profile of the auto-covariance function
$a(s):=\langle H(h_{k}(t+s)-\theta)\,H(h_{k}(t)-\theta)\rangle\simeq\langle m\rangle+\langle m\rangle(1-\langle m_{k}\rangle)\,e^{-\frac{|s|}{\tau}}$,
which has the right asymptotic behaviors ($a(0)=m_{k}$ and $a(s\to\infty)=m_{k}^{2}$)
and assumes fluctuations to be correlated on the time scale $\tau$
of the update, leads to the approximate value $\langle H(h_{k}(t)-\theta)\,n_{k}(t)\rangle\simeq\frac{1}{2}m_{k}(1-m_{k})$
\citep[by using ][ eq. (A.10)]{Vreeswijk98}. The corresponding approximation
of the slope following from the former approximation with \prettyref{eq:def_D}
is shown in \prettyref{fig:Reconstruction-of-connectivity}a. However,
the resulting expression for the slope at zero time lag cannot be
written in the form \prettyref{eq:q_kl}. Therefore, we assume that
\prettyref{eq:q_kl} also holds on the diagonal, although the resulting
predicted slopes have a slight negative bias compared to simulation
results.

Given the two covariance matrices entering \prettyref{eq:def_D},
as well as the average activities $m_{i}$ for each neuron, we can
derive a procedure for reconstructing the connectivity $J_{kl}/\sigma_{k}$.
The mean and variance of the input to a neuron are determined by \prettyref{eq:mu_sigma}.
Inverting the relation \prettyref{eq:mean_erfc} as $y(m_{k}):=\frac{\mu_{k}-\theta}{\sqrt{2}\sigma_{k}}=-\erfc^{-1}(2m_{k})$
allows us to determine the susceptibility \prettyref{eq:susceptibility-1}
as $S\left(m_{k},\sigma_{k}\right)=\frac{1}{\sqrt{2\pi}\sigma_{k}}\,e^{-y(m_{k})^{2}}$.
Hence we obtain an expression for the ratio of the incoming synaptic
weight and the total synaptic noise of neuron $k$ 
\begin{eqnarray}
\frac{J_{kl}}{\sigma_{k}}= & \sqrt{2\pi}w_{kl}e^{y(m_{k})^{2}} & .\label{eq:reconstructed_weight}
\end{eqnarray}
This again shows that correlations are only controlled by the ratio
$J_{kl}/\sigma_{k}$ rather than by $J_{kl}$ and $\sigma_{k}$ alone,
in line with the invariance found in \prettyref{sec:scale_invariance}.
\prettyref{fig:Reconstruction-of-connectivity}b shows the reconstruction
of connectivity from the simulated covariance functions. Due to the
previously discussed approximations, the exact reconstruction of the
relative weights $J_{kl}/\sigma_{k}$ is not possible for a binary
network. The observed deviations from the identity line are predominantly
caused by the approximation of the slope of the auto-covariance functions.
The reconstructed connectivity correctly infers all excitatory and
all inhibitory connections, but additionally yields a considerable
number of false positive excitatory and inhibitory connections.

The described procedure, moreover, allows us to determine the remaining
parameters of the binary network. With regard to correlations, we
are free to choose an arbitrary $\sigma_{k}$, e.g. $\sigma_{k}=1$
to determine $J_{kl}$ by \eqref{eq:reconstructed_weight}. As the
mean activity and correlations are known, the actual magnitude of
fluctuations $\sigma_{k}^{\mathrm{loc}}$ caused by the local inputs
from the network follows from \prettyref{eq:mu_sigma}. If these fluctuations
are smaller than or equal to our arbitrary choice ($\sigma_{k}^{\mathrm{loc}}\le\sigma_{k}=1$),
it is possible to supply each neuron with additional noise of variance
$1-\left(\sigma_{k}^{\mathrm{loc.}}\right)^{2}$. Only in this case
we can construct a binary network satisfying the given constraints.
Having fixed $J_{kl}$, and given the mean activities by \prettyref{eq:mu_sigma},
determines the mean input $\mu_{k}$ to each cell. Finally the threshold
$\theta$ must be chosen such that $\theta_{k}=\mu_{k}-\sqrt{2}y(m_{k})$.

Related methods for systems with sequential Glauber update in discrete
time steps have previously been proposed. \citet{Mezard11_L07001}
present a method that relates the equal time and one step delayed
covariance matrices to the connectivity (see their eqs. (8) and (18)).
A similar approach, termed TAP, is described in \citep[see their eq. (14)]{Zheng11_041135}.
In contrast, in the time-continuous case considered here, the sum
of the covariance function and its slope appear on the left hand side
of \eqref{eq:def_D}. Qualitatively, the results for discrete and
for continuous dynamics agree, as the latter sum presents the first
order Taylor approximation of the covariance function at a time lag
$\tau$, which is equal to the average update interval.

\section{Discussion\label{sec:Discussion}}

We here present a theoretical description of fluctuations in strongly
coupled networks of large numbers of binary units that, for the first
time, faithfully captures the statistics of individual units as well
as pairs of units. The fluctuations are characterized by self-consistent
equations for mean activity and pairwise correlations, including finite-size
effects down to hundreds of units. The method can be applied to a
wide range of networks, in particular networks with asymmetric and
strong couplings.

Standard approaches describing symmetric systems in terms of a partition
function cannot be extended to non-symmetric coupling, because these
systems do not reach thermodynamic equilibrium. Here, we set out to
capture the statistics of the system in terms of the cumulant hierarchy
for the activity variables. Truncating the hierarchy after second
order yields a closed set of equations that already provides a good
approximation for the zero time lag covariances between individual
units. We show the equivalence of this truncation to the Gaussian
approximation of the input field \citep{Vreeswijk98,Renart10_587},
which follows from the central limit theorem. Non-Gaussian, finite-size
effects in networks of only several hundreds of units are effectively
taken into account by incorporating a subset of third order cumulants
that can be expressed in terms of lower order cumulants due to the
binary nature of the activity variables. This significantly improves
the prediction of mean activities, which are strongly affected by
the non-linearity of the gain function. The resulting set of non-linear
coupled equations can efficiently be solved numerically by damped
fixed-point iteration.

We demonstrate that a Heaviside activation function - independent
of the choice of coupling amplitudes - constitutes the strongest possible
interaction for binary networks. This finding generalizes the invariance
of population-averaged pairwise correlations under proportional scaling
of couplings and threshold of all units \citep{Grytskyy13_258,Albada15}
to the invariance of individual pairwise correlations under the scaling
of the corresponding units' parameters. Weaker coupling in binary
networks can only be achieved by a smoother activation function, but
not by different scalings of coupling amplitudes, as e.g. $1/N$ versus
$1/\sqrt{N}$, contradictory to frequently employed arguments in the
literature \citep[as discussed in][]{Albada15}. While the scaling
invariance is generic, the presented Gaussian and close-to-Gaussian
approximations of the input field hold for the commonly considered
narrowly distributed couplings. For wide distributions, the statistics
of the input field is likely to depart from the close-to Gaussian
assumption.

The contribution of cross-covariances to the marginal statistics of
the input to each neuron causes a distribution of the mean activities,
even in networks composed of neurons each receiving an identical number
of inputs. The classical treatment of neurons with a Heaviside non-linearity
neglects this effect \citep{Vreeswijk98}. Still, distributed numbers
of synaptic inputs typically dominate the distribution of mean activities
in sparsely \citep{Vreeswijk98} and densely connected random networks
\citep{Renart10_587,Helias14}.

The Gaussian closure, i.e. the truncation of the cumulant hierarchy
after second order, yields a set of equations that accurately predicts
the second order statistics. This is in line with experimental evidence
showing that pairwise correlations sufficiently constrain maximum
entropy models of collective activity \citep{Schneidman06_1007}.
Analogously, the truncation after first order, i.e. the Curie-Weiss
mean-field theory that neglects fluctuations altogether, already yields
a good estimate of the first moments. In the context of balanced networks
this description is sometimes referred to as ``balance equations''
\citep[ eqs. (4.1) and (4.2)]{Vreeswijk98}. Taking into account fluctuations
in the input, due to the variance of individual units \citep{Vreeswijk98},
constitutes an intermediate step in the cumulant hierarchy. The variance
can be determined exploiting the binary character of the variables
resulting in all cumulants of a single unit being fixed by the mean.
A consequence of the lower order moments depending only weakly on
the statistics of higher order moments is that cross-covariances can
be determined from linear fluctuations around the steady state that
itself is determined neglecting correlations altogether. We may therefore
speculate that an approximation of third or higher order correlations
can be obtained from the fluctuations around a state that itself is
determined self-consistently by taking into account only up to second
order correlations.

The consistent truncation of the cumulant hierarchy further provides
deeper insights: The expression for covariances between individual
neuron pairs follows naturally from a consistent Gaussian closure,
without further approximation. In previous studies, it was obtained
as a linear approximation for weak correlations \citep[supplementary eqs. (31)-(33)]{Renart10_587}
and population-averaged covariances \citep[eq. (6.8)]{Ginzburg94}.
We here show that solving the modified Lyapunov equation naturally
leads to a decomposition of fluctuations into eigenmodes of the system.
The presented method solves the problem that the modified Lyapunov
equation does not hold on the diagonal, in contrast to the case of
population averages \citep{Ginzburg94}. The approach moreover exposes
that fluctuations in the Gaussian approximation are described by a
set of coupled Ornstein-Uhlenbeck processes. It is well known that
the fluctuations obtained from a systematic system-size expansion
\citep[chapter X]{VanKampen1992_StochasticProcesses}, to lowest order,
obey a Langevin equation. This result has been applied to networks
of homogeneous populations \citep{Ohira95_290,Bressloff09_1488,Grytskyy13_131}.
The interesting point here is the feasibility of such a reduction
directly on the level of individual binary variables. In contrast
to the population-averaged activity, the jumps of a binary variable
cannot be considered small compared to its value, hence the precondition
to apply the system-size expansion is not valid. The approach taken
here to expose the correspondence between binary dynamics and Ornstein-Uhlenbeck
processes is hence complementary. In particular, the result shows
that the effective noises appearing in the equivalent set of Ornstein-Uhlenbeck
processes are not uniquely determined by the activities of the neurons
alone, in contrast to the population-averaged case \citep[sec. 3.1]{Grytskyy13_131}. 

Decomposing externally applied signals to the network into the same
eigenmodes as the intrinsically generated fluctuations, we obtain
an expression for the susceptibility of the network on the single
unit level that explains the spatio-temporal filtering applied to
external signals, complementing the result for interacting populations
\citep[sec. V]{Ginzburg94}. The expression shows how the decay properties
of the induced network response are determined by the eigenvalue associated
to the eigendirections stimulated by the external signal. In particular,
stimuli exciting modes close to an instability yield responses with
a long memory life time. In Erd\H{o}s-R\'enyi random binary networks
with fixed weights, the spectral radius of the linearized connectivity
matrix is bounded by $\sqrt{2(1-p)/\pi}<0.8<1$ \citep{Grytskyy13_258},
so that all modes of the network are stable. An instability of the
mean activities can only be achieved with connectivity statistics
or motifs that differ from an Erd\H{o}s-R\'enyi network. This finding
is in qualitative contrast to spiking networks that indeed show a
rate instability at a critical coupling weight \citep{Ostojic14,Harish15_e1004266}.
This insight has consequences for the use of random networks with
fixed weights in the framework of reservoir computing \citep{Jaeger01_echo,Maass02_2531},
where highest computational performance \citep{Crutchfield89_105}
is achieved at the edge of chaos \citep[for a review, see][]{Legenstein07_127}.

We show that the effective interaction strength, i.e. the product
of a unit's susceptibility and the incoming coupling amplitude, can
uniquely be reconstructed from the covariance matrix and the slope
of the covariance functions at zero time lag. This relation builds
on the regression theorem \citep{Risken96}, which states that the
fluctuations in the system, to linear order \eqref{eq:OUP_fluct},
follow the same differential equation as the time-delayed covariance
functions \eqref{eq:1st_2nd_moment}. Coupling amplitudes, for principle
reasons, cannot be inferred unambiguously: Correlations only depend
on $J/\sigma$, where $\sigma$ measures the strength of the overall
incoming fluctuations to the node. The inference of the effective
interaction strength $J/\sigma$ still allows the classification of
connections into excitatory and inhibitory ones, but leads to a considerable
rate of false positive connections. The method presented here requires
the measurement of covariances between all pairs of units in the network.
In case of severe undersampling, its application is restricted to
small sub-networks. Thus, in general settings, more sophisticated
approaches that take into account the influence of hidden units on
the observed covariances, as e.g. in \citet{Tyrcha13_1301} are more
promising. The presented expressions that relate the covariance matrix
and its slope at zero time lag to the connectivity may still proof
useful to construct similar inference methods for the investigated
class of networks. Moreover, the algorithm constructing a network
that generates activity with desired mean activities and covariances
is a useful tool to generate surrogate data.

Previous works have addressed several aspects of pairwise correlations.
The smallness of the average magnitude of covariances in strongly
coupled balanced networks \citep{Vreeswijk96} has been explained
by the influential work of \citet{Renart10_587} in the large $N$
limit. In general, decorrelation follows from the dominant negative
feedback in balanced networks at any network size \citep{Tetzlaff12_e1002596,Helias14}.
These global properties are determined by collective fluctuations
and can hence be described by population-averaged mean-field theory
\citep{Ginzburg94}. Here we discuss the statistics of individual
units, which therefore goes beyond the existing approaches.

Our results are complementary to the method presented in \citet{Buice09_377},
who consider Markov systems of interacting populations of neurons
that consequently have integer numbers of active states, as opposed
to the Glauber dynamics \citep{Glauber63_294} of individual neurons
considered here. In consequence, the truncation in the former work
is performed on the level of normal-ordered cumulants, that represent
an approximation around Poisson statistics, while here we derive an
expansion in terms of ordinary cumulants \citep{Gardiner04}. Moreover,
\citet{Buice09_377} rely on a smooth activation function, while our
results are generally applicable, including deterministic single units
with hard thresholds. Both formalisms generalize to higher order cumulants
of activity variables.

A different approach treating the statistics of individual units in
a single network realization analytically originates in the spin glass
literature. Initiated by the symmetrically coupled Sherrington-Kirkpatrick
spin glass model \citep{Sherrington75_1792}, an expansion of the
free energy in the coupling strength (Plefka-expansion \citep{Plefka82_1971})
leads to the Thouless-Anderson-Palmer mean-field theory \citep{Thouless77_593,Nakanishi97_8085,Tanaka98_2302,Nishimori01_01}.
By construction this method is restricted to weak coupling and, in
its original form, starting from the partition function, also to systems
in thermodynamic equilibrium. An extension to asymmetrically coupled
non-equilibrium systems has been derived invoking information theoretic
arguments \citep{Kappen00_5658}. A related approach has been taken
by \citep{Zheng11_041135}. Still, the extended methods rely on the
smoothness of the activation function and the smallness of the coupling
constants, two restrictions that we are able to overcome here. Compared
to previously mentioned works, the presented approach is closer related
to the mean-field theory by \citet{Mezard11_L07001}. Although they
consider a system with sequential Glauber update in discrete time
steps, the method can be extended to the asynchronous update investigated
here. Their analysis does not require weak coupling, similar to ours,
but in contrast still relies on a smooth activation function for individual
units. Moreover, their theory neglects the influence of the cross-covariance
on the marginal statistics (their eq. (4)), an important finite-size
effect here shown to result in a distribution of mean activities due
to correlations alone.

The mean-field methods employed in computer science, biology, artificial
intelligence, social sciences, economics, and theoretical neuroscience
\citep[see e.g. ][]{Fernandez14_158701,Aljadeff15_088101,Kadmon15_041030}
may be complemented by our results that go beyond population-averaged
dynamics. The general formalism starting from the master equation
can be widely adapted by defining model-specific transition rates
\citep[see Tab. 1 in ][]{Gleeson13_021004}. In neuroscience, with
the advancement of electrophysiology, experimental data with hundreds
of simultaneously recorded neurons has become readily available \citep{Buzsaki04_446,Brown04_456}.
The availability of parallel data progressively changes the focus
in neuroscience from the study of single cell responses to emergent
phenomena arising through the interaction between neurons in networks
\citep{Stevenson11_139}. Pairwise correlations are moreover closely
linked to fluctuations of the population activity \citep{Tetzlaff12_e1002596},
which have been shown to shape experimentally accessible signals,
such as the local field potential or the electroencephalogram (EEG)
\citep{Linden11_859,Mazzoni10_956}. The effective description of
the statistics of individual neurons, valid in the entire range of
coupling strength, forms the basis for studies of dynamics on adaptive
networks \citep{Gross06_208701}, e.g. the interaction of neuronal
dynamics with correlation-sensitive learning rules \citep{Ocker15_e1004458}.
Explicit expressions not only for zero time lag, but also for the
slope of covariance functions allow the definition of plasticity rules
resembling spike-timing dependent plasticity \citep{Morrison08_459}.
The finding that the collective dynamics is captured by the non-trivial
second order statistics implies that theoretical descriptions of mechanisms,
e.g. biologically realistic synaptic learning, which usually only
rely on first and second order statistics, have now come into reach.
Balanced networks show widely distributed correlations across pairs
of neurons with small mean \citep{Ecker10}. This robust feature is
captured by the presented linear equations for the covariance matrix.
The shape of the distribution has to date not been related to the
properties of the underlying network structure. Further work is required
to obtain analytical expressions exposing how the structural properties
give rise to the statistics of the distribution of covariances. Such
results would enable us to deduce statistics of the connections from
the observed activity. The presented description of the structure
of fluctuations in this archetypical model of collective phenomena
by a set of non-linear equations provides a starting point in this
endeavor and facilitates further development on disordered, coupled
systems with large numbers of degrees of freedom.

\section{Appendix}

\subsection{Derivation of moment equations up to second order\label{sub:Derivation-of-moment-equations}}

For completeness and to establish a consistent notation, we here include
the derivation of equations \prettyref{eq:1st_2nd_moment} for the
first and second moments of the activity in a binary network. We follow
the notation introduced in \citet{Buice09_377} and used in \citet{Helias14}.

The stochastic system is completely characterized by the joint probability
distribution $p(\n)$ of all $N$ binary variables $\n$. Knowing
the joint probability distribution, arbitrary moments can be calculated,
among them pairwise correlations. The occupation probability of each
state follows the master equation \prettyref{eq:balancing_flux}.
We denote as $\n_{i+}=(n_{1},\ldots,n_{i-1},1,n_{i+1},\ldots,n_{N})$
the state, where the $i$-th neuron is active ($n_{i}=1$), and $\n_{i-}$
where neuron $i$ is inactive ($n_{i}=0$). Since in each infinitesimal
time interval at most one neuron can change state, for each given
state $\n$ there are $N$ possible transitions (each corresponding
to one of the $N$ neurons changing state). The sum of the probability
fluxes into the state and out of the state must sum up to the change
of probability in the respective state \citep{Kelly79}, i.e.

\begin{align}
\tau\frac{\partial p(\n)}{\partial t}= & \sum_{i=1}^{N}(2n_{i}-1)\left(p(\n_{i-})F_{i}(\n_{i-})\right.\nonumber \\
 & \hphantom{\sum_{i=1}^{N}(2n_{i}-1)}-\left.p(\n_{i+})(1-F_{i}(\n_{i+})\right)\label{eq:balancing_flux}
\end{align}
$\forall\quad\n\in\{0,1\}^{N}.$ From this equation we derive expressions
for the first $\langle n_{k}\rangle$ and second moments $\langle n_{k}n_{l}\rangle$
by multiplying with $n_{k}n_{l}$ and summing over all possible states
$\n\in\{0,1\}^{N}$, which leads to
\begin{eqnarray*}
\tau\frac{\partial}{\partial t}\langle n_{k}n_{l}\rangle & = & \sum_{\n\in\{0,1\}^{N}}\sum_{i=1}^{N}n_{k}n_{l}(2n_{i}-1)\\
 &  & \times\underbrace{\left(p(\n_{i-})F_{i}(\n_{i-})-p(\n_{i+})(1-F_{i}(\n_{i+}))\right)}_{\equiv G_{i}(\n\backslash n_{i})},
\end{eqnarray*}
where we denote as $\langle f(\n)\rangle=\sum_{\n\in\{0,1\}^{N}}p(\n)f(\n)$
the average of a function $f(\n)$ with respect to the distribution
$p(\n)$. Note that the term denoted $G_{i}(\n\backslash n_{i})$
does not depend on the state of neuron $i$. We use the notation $\n\backslash n_{i}$
for the state of the network excluding neuron $i$, i.e. $\n\backslash n_{i}=(n_{1},\ldots,n_{i-1},n_{i+1},\ldots,n_{N})$.
Separating the terms in the sum over $i$ into those with $i\neq k,l$
and the two terms with $i=k$ and $i=l$, we obtain

\begin{eqnarray*}
\tau\frac{\partial}{\partial t}\langle n_{k}n_{l}\rangle & = & \sum_{\n}\;\sum_{i=1,i\neq k,l}^{N}n_{k}n_{l}(2n_{i}-1)\,G_{i}(\n\backslash n_{i})\\
 &  & \hphantom{\sum_{\n}\;\sum_{i=1,i\neq k,l}^{N}}+n_{k}n_{l}(2n_{k}-1)\,G_{k}(\n\backslash n_{k})\\
 &  & \hphantom{\sum_{\n}\;\sum_{i=1,i\neq k,l}^{N}}+n_{k}n_{l}(2n_{l}-1)\,G_{l}(\n\backslash n_{l})\\
 & = & \sum_{i=1,i\neq k,l}^{N}\;\sum_{\n\backslash n_{i}}n_{k}n_{l}(G_{i}(\n\backslash n_{i})-G_{i}(\n\backslash n_{i}))\\
 &  & +\sum_{\n}n_{k}n_{l}\,G_{k}(\n\backslash n_{k})+\sum_{\n}n_{k}n_{l}\,G_{l}(\n\backslash n_{l}),
\end{eqnarray*}
where we obtained the first term by explicitly summing over state
$n_{i}\in\{0,1\}$ (i.e. using $\sum_{\n\in\{0,1\}^{N}}=\sum_{\n\backslash n_{i}\in\{0,1\}^{N-1}}\sum_{n_{i}=0}^{1}$
and evaluating the sum $\sum_{n_{1}=0}^{1}$). This first sum obviously
vanishes. The remaining terms are of identical form with the roles
of $k$ and $l$ interchanged. We hence only consider the first of
them and obtain the other by symmetry. The first term simplifies to
\begin{eqnarray}
 &  & \sum_{\n}n_{k}n_{l}\,G_{k}(\n\backslash n_{k})\nonumber \\
 & \stackrel{\text{\ensuremath{{\scriptstyle \mbox{\ensuremath{n_{k}=1}}}}}}{=} & \sum_{\n\backslash n_{k}}n_{l}\,G_{k}(\n\backslash n_{k})\nonumber \\
 & \stackrel{\text{\ensuremath{{\scriptstyle \text{def. }\mbox{\ensuremath{G_{k}}}}}}}{=} & \begin{cases}
\sum_{\n\backslash n_{k}}p(\n_{k-})\,F_{k}(\n_{k-})\\
\hphantom{\sum_{\n\backslash n_{k}}}+p(\n_{k+})\,F_{k}(\n_{k+})-p(\n_{k+}) & \text{for }k=l\\
\sum_{\n\backslash n_{k}}p(\n_{k-})\,n_{l}\,F_{k}(\n_{k-})\\
\hphantom{\sum_{\n\backslash n_{k}}}+p(\n_{k+})\,n_{l}\,F_{k}(\n_{k+})-n_{l}\,p(\n_{k+}) & \text{for }k\neq l
\end{cases}\nonumber \\
 & = & \begin{cases}
\langle F_{k}(\n)\rangle-\langle n_{k}\rangle & \text{for }k=l\\
\langle F_{k}(\n)\,n_{l}\rangle-\langle n_{k}n_{l}\rangle & \text{for }k\neq l
\end{cases}.\label{eq:intermediate_G}
\end{eqnarray}
Taken together with the mirror term $k\leftrightarrow l$, we arrive
at two conditions, one for the first ($k=l$, $\langle n_{k}^{2}\rangle=\langle n_{k}\rangle$)
and one for the second ($k\neq l$) moment
\begin{align}
\tau\frac{\partial}{\partial t}\langle n_{k}\rangle & \stackrel{\text{\ensuremath{{\scriptstyle \mbox{\ensuremath{k=l}}}}}}{=}-\langle n_{k}\rangle+\langle F_{k}(\n)\rangle\label{eq:mean_activity}\\
\tau\frac{\partial}{\partial t}\langle n_{k}n_{l}\rangle & \stackrel{\text{\ensuremath{{\scriptstyle \mbox{\ensuremath{k\neq l}}}}}}{=}-2\langle n_{k}n_{l}\rangle+\langle F_{k}(\n)\,n_{l}\rangle+\langle F_{l}(\n)\,n_{k}\rangle.\label{eq:second_moment_diffeq}
\end{align}
The time-lagged correlation function can be derived along completely
analogous lines as \prettyref{eq:second_moment_diffeq}, as the forward
time evolution equation (differential equation with respect to $t$)
of the two point probability distribution $p(\n,t,\q,s)$ fulfills,
due to the Markov property, the same master equation \prettyref{eq:balancing_flux}
as the equal time probability distribution $p(\n,t)$. The resulting
differential equation reads
\begin{align}
\tau\frac{\partial}{\partial t}\langle n_{k}(t)n_{l}(s)\rangle & \equiv\sum_{\n,\q}p(\n,t,\q,s)\,n_{k}q_{l}\nonumber \\
 & =\ldots\nonumber \\
 & =-\langle n_{k}(t)n_{l}(s)\rangle+\langle F_{k}(\n(t))\,n_{l}(s)\rangle.\label{eq:time_lagged_correlation}
\end{align}

\subsection{Cumulants of summed random variables}

Let 
\begin{eqnarray}
y & = & \sum_{i=1}^{N}x_{i},\label{eq:def_y}
\end{eqnarray}
with $x_{i}$ random variables that follow an arbitrary distribution
$p(\x)$. The moment generating functions of $y$ and $\x$ are then
related by
\begin{eqnarray}
\varphi_{y}(t) & = & \sum_{\x}p(\x)e^{ty}=\langle e^{ty}\rangle_{\x}\label{eq:def_varphi_yx}\\
 & = & \langle e^{t\sum_{i=1}^{N}x_{i}}\rangle_{\x}\nonumber \\
 & = & \left(\varphi_{\x}\circ\iota\right)(t),\nonumber 
\end{eqnarray}
where concatenation with the function $\iota:t\mapsto(t,\ldots,t)$
replaces every $t_{i}$ by a $t$. The cumulant generating functions
therefore follow as
\begin{eqnarray}
\Phi_{y}(t) & = & \ln\varphi_{y}(t)\nonumber \\
 & = & \ln\left(\varphi_{\x}\circ\iota\right)(t)=\left(\Phi_{\x}\circ\iota\right)(t),\label{eq:def_Phi_y}
\end{eqnarray}
with $\Phi_{\x}(\t)=\ln\langle e^{\t\cdot\x}\rangle$ the cumulant
generating function of the $\x$ and $\t\cdot\x$ the scalar product.
From the expansion in cumulants $\kappa_{ij\ldots}^{x}$ for $\x$
and $\kappa_{1,2,\ldots}$ for $y$ follows the relationship
\begin{eqnarray}
\Phi_{y}(t) & \equiv & \sum_{l=1}^{\infty}\frac{\kappa_{l}}{l!}t^{l}\nonumber \\
 & = & \left(\Phi_{\x}\circ\iota\right)(t)\nonumber \\
 & = & \left[\left(\sum_{i=1}^{N}\kappa_{i}^{x}t_{i}+\sum_{i,j=1}^{N}\frac{\kappa_{ij}^{x}}{2!}\,t_{i}t_{j}+\ldots\right)\circ\iota\right](t)\nonumber \\
 & = & \underbrace{\sum_{i=1}^{N}\kappa_{i}^{x}}_{\kappa_{1}}\,t+\frac{1}{2!}\underbrace{\sum_{i,j=1}^{N}\kappa_{ij}^{x}}_{\kappa_{2}}\,t^{2}+\ldots,\label{eq:cumulant_sum}
\end{eqnarray}
so the cumulants of the summed variable $\kappa_{1}$, $\kappa_{2}$,
etc are given by the sums of the cumulants of the individual variables
of corresponding order.

\subsection{Functions of close-to Gaussian variables\label{sub:Functions-close-to-Gaussian}}

To determine the mean activity, we need to apply a non-linear function
$f$ to a variable $h_{k}$ that has a statistics which is close to
Gaussian. For brevity, we will suppress the index $k$ in the following.
We assume that the Fourier transform $\hat{f}(\omega)$ exists. Let
$y$ be the random variable \eqref{eq:def_y} which is a sum of a
large number of individual variables $x_{i}$ and is assumed to be
close to Gaussian. For the expectation value $\langle f(y)\rangle_{y}$
we get 
\begin{align}
\langle f(y)\rangle_{\x} & =\frac{1}{2\pi}\int d\omega\,\hat{f}(\omega)\,\langle e^{i\omega y}\rangle_{\x}\nonumber \\
 & =\frac{1}{2\pi}\int d\omega\,\hat{f}(\omega)\,e^{\Phi_{y}(i\omega)}\nonumber \\
 & =\frac{1}{2\pi}\int d\omega\,\hat{f}(\omega)\,e^{\kappa_{1}(i\omega)+\frac{1}{2!}\kappa_{2}(i\omega)^{2}+\frac{1}{3!}\kappa_{3}(i\omega)^{3}+\ldots},\label{eq:Fourier_moment_gen}
\end{align}
where $\Phi_{y}$ is the cumulant generating function of $y$ \eqref{eq:def_Phi_y}
and $\kappa_{i}$ denotes the $i$-th cumulant of $y$. We are interested
in an approximation that treats the dominant first and second (Gaussian)
cumulants of $y$ explicitly and separate the effect of all higher
cumulants by writing 
\begin{eqnarray}
\langle f(y)\rangle_{\x} & = & e^{\frac{1}{6}\kappa_{3}\left(\frac{\partial}{\partial\kappa_{1}}\right)^{3}+\ldots}\int d\omega\,\hat{f}(\omega)\,e^{\kappa_{1}(i\omega)+\frac{1}{2}\kappa_{2}(i\omega)^{2}}\nonumber \\
 & = & e^{\frac{1}{6}\kappa_{3}\left(\frac{\partial}{\partial\kappa_{1}}\right)^{3}+\ldots}\langle f(y)\rangle_{y\sim\mathcal{N}(\kappa_{1},\kappa_{2})},\label{eq:<F(y)>_3rd}
\end{eqnarray}
where we identified in the last line the Fourier transform of a Gaussian
with moments $\kappa_{1}$ and $\kappa_{2}$ via \eqref{eq:Fourier_moment_gen}.

For the covariance we need to evaluate terms of the form $\langle F_{k}n_{l}\rangle$.
These terms can be obtained analogously as

\begin{eqnarray*}
\langle f(y)x_{l}\rangle_{\x} & = & \frac{1}{2\pi}\int d\omega\,\hat{f}(\omega)\,\langle x_{l}\,e^{i\omega y}\rangle_{\x}\\
 & = & \frac{1}{2\pi}\int d\omega\,\hat{f}(\omega)\,\left(\left(\partial_{t_{l}}\varphi_{\x}\right)\circ\iota\right)(i\omega)\\
 & = & \frac{1}{2\pi}\int d\omega\,\hat{f}(\omega)\,\left(\left(\left[\partial_{t_{l}}\Phi_{\x}\right]\,e^{\Phi_{\x}}\right)\circ\iota\right)(i\omega),
\end{eqnarray*}
where $\varphi_{\x}$ is the moment generating function of $\x$ \prettyref{eq:def_varphi_yx}.
The derivative of the cumulant generating function with \eqref{eq:cumulant_sum}
becomes 
\begin{eqnarray*}
\partial_{t_{l}}\Phi_{\x}(\t) & = & \kappa_{l}^{x}+\sum_{j=1}^{N}\kappa_{lj}^{x}\,t_{j}+\frac{1}{2}\sum_{i,j=1}^{N}\kappa_{lij}^{x}\,t_{i}t_{j}+\ldots\\
\left(\partial_{t_{l}}\Phi_{\x}\circ\iota\right)(i\omega) & = & \kappa_{l}^{x}+\underbrace{\sum_{j=1}^{N}\kappa_{lj}^{x}}_{=:\Delta\kappa_{1,l}}\,(i\omega)+\frac{1}{2}\underbrace{\sum_{i,j=1}^{N}\kappa_{lij}^{x}}_{=:\Delta\kappa_{2,l}}\,(i\omega)^{2}+\ldots\\
 & = & \sum_{q=0}^{\infty}\Delta\kappa_{q,l}(i\omega)^{q},
\end{eqnarray*}
where product rule produced a factor $2$ in the second term and a
factor $3$ in the third term and we used the symmetry of the cumulants
with respect to permutations of indices. So we get

\begin{eqnarray}
\langle x_{l}f(y)\rangle_{\x} & = & \int d\omega\,\hat{f}(\omega)\,\sum_{q=0}^{\infty}\Delta\kappa_{q,l}(i\omega)^{q}e^{\sum_{p=0}^{\infty}\kappa_{p}(i\omega)^{p}}\nonumber \\
 & = & \left[\sum_{q=0}^{\infty}\Delta\kappa_{q,l}\,\frac{\partial}{\partial\kappa_{q}}\right]\,e^{\frac{1}{3!}\kappa_{3}\left(\frac{\partial}{\partial\kappa_{1}}\right)^{3}+\ldots}\langle f(y)\rangle_{y\sim\mathcal{N}(\kappa_{1},\kappa_{2})}\nonumber \\
 & = & \left[\sum_{q=0}^{\infty}\Delta\kappa_{q,l}\,\frac{1}{q!}\,\left(\frac{\partial}{\partial\kappa_{1}}\right)^{q}\right]\,\langle f(y)\rangle_{\x},\label{eq:<fx>}
\end{eqnarray}
where we replaced $(i\omega)^{q}$ by derivatives of the exponential
with respect to cumulants and identified $\langle f(y)\rangle_{\x}$
in the last step using \eqref{eq:<F(y)>_3rd}.

\subsection{Trivial third and fourth order cumulants of binary variables expressed
by lower order cumulants\label{sub:Trivial-third-order}}

\textbf{Third order.} Let $n_{l},n_{i},n_{j},n_{r}\in[0,1]$ be binary
variables. The raw third moment can be written as a sum of all combinations
of cumulants up to order three \prettyref{eq:third_moment} $\langle n_{l}n_{i}n_{j}\rangle=\llangle n_{l}n_{i}n_{j}\rrangle+c_{li}m_{j}+c_{ij}m_{l}+c_{jl}m_{i}+m_{l}m_{i}m_{j}.$
Using $n_{i}^{K}=n_{i}$ for each integer $K\ge1$, in case of binary
variables $n_{i}$ we consider the two cases

\begin{eqnarray}
l=i\neq j:\quad\langle n_{l}n_{j}\rangle & = & \llangle n_{l}n_{l}n_{j}\rrangle\nonumber \\
 & + & c_{ll}m_{j}+c_{lj}m_{l}+c_{jl}m_{l}\nonumber \\
 & + & m_{l}^{2}m_{j}\nonumber \\
\llangle n_{l}n_{l}n_{j}\rrangle & = & \underbrace{\langle n_{l}n_{j}\rangle}_{c_{lj}+\langle n_{l}\rangle\langle n_{j}\rangle}-c_{ll}m_{j}-2c_{lj}m_{l}-m_{l}^{2}m_{j}\nonumber \\
 & = & c_{lj}(1-2m_{l})+(\underbrace{-c_{ll}+m_{l}-m_{l}^{2}}_{=0})m_{j}\nonumber \\
 & = & c_{lj}(1-2m_{l})\nonumber \\
\nonumber \\
l=i=j:\quad\langle n_{l}\rangle & = & \llangle n_{l}n_{l}n_{l}\rrangle+3c_{ll}m_{l}+m_{l}^{3}\nonumber \\
\llangle n_{l}n_{l}n_{l}\rrangle & = & m_{l}-3m_{l}(1-m_{l})m_{l}-m_{l}^{3}\nonumber \\
 & = & m_{l}-3m_{l}^{2}+2m_{l}^{3}\nonumber \\
 & = & c_{ll}(1-2m_{l}),\label{eq:kappa_kkl_kkk}
\end{eqnarray}
which together yield the expression \eqref{eq:kappa_llj} in the
main text.

The third cumulant \prettyref{eq:3rd_cumulant_input} of $h_{k}$
follows with the assumption $\llangle n_{i}n_{j}n_{r}\rrangle\simeq0$
for $i\neq j\neq r$ as

\begin{eqnarray}
\kappa_{3,k} & = & \sum_{ijr}J_{ki}J_{kj}J_{kr}\llangle n_{i}n_{j}n_{r}\rrangle\nonumber \\
 & = & \underbrace{\sum_{i\neq j\neq r}}_{N(N-1)(N-2)\text{ terms}}J_{ki}J_{kj}J_{kr}\underbrace{\llangle n_{i}n_{j}n_{r}\rrangle}_{\simeq0}\nonumber \\
 &  & +3\underbrace{\sum_{i=j\neq r}}_{N(N-1)\text{ terms}}J_{ki}^{2}J_{kr}\llangle n_{i}n_{i}n_{r}\rrangle+\underbrace{\sum_{i=j=r}}_{N\text{ terms}}J_{ki}^{3}\llangle n_{i}n_{i}n_{i}\rrangle\nonumber \\
 & \simeq & 3\sum_{i=j\neq r}J_{ki}^{2}J_{kr}c_{ir}(1-2m_{i})+\sum_{i=j=r}J_{ki}^{3}c_{ii}(1-2m_{i})\nonumber \\
 & = & 3\sum_{i,r}J_{ki}^{2}(1-2m_{i})c_{ir}J_{kr}-2\sum_{i}J_{ki}^{3}c_{ii}(1-2m_{i}),\label{eq:lambda_k_app}
\end{eqnarray}
where we included the term $i=r$ in the first sum and compensated
accordingly in the latter term. With $c_{ii}(1-2m_{i})=m_{i}-3m_{i}^{2}+2m_{i}^{3}$
we obtain the expression \prettyref{eq:lambda_k} in the main text.
Analogously follows

\begin{eqnarray}
 &  & \sum_{i,j=1}^{N}J_{ki}J_{kj}\llangle n_{l}n_{i}n_{j}\rrangle\nonumber \\
 & = & \underbrace{\sum_{i\neq j\neq l}J_{ki}J_{kj}\llangle n_{l}n_{i}n_{j}\rrangle}_{\simeq0}+\sum_{l\neq i=j=1}^{N}J_{ki}^{2}\llangle n_{l}n_{i}n_{i}\rrangle\nonumber \\
 &  & +\sum_{i=l\neq j=1}^{N}J_{kl}J_{kj}\llangle n_{l}n_{l}n_{j}\rrangle+\sum_{j=l\neq i=1}^{N}J_{ki}J_{kl}\llangle n_{l}n_{i}n_{l}\rrangle\nonumber \\
 &  & +J_{kl}^{2}\llangle n_{l}n_{l}n_{l}\rrangle\nonumber \\
 & = & \sum_{l\neq i=1}^{N}J_{ki}^{2}\,c_{li}(1-2m_{i})+2J_{kl}\,\sum_{l\neq i=1}^{N}J_{ki}c_{li}(1-2m_{l})\nonumber \\
 &  & +J_{kl}^{2}c_{ll}(1-2m_{l})\nonumber \\
 & = & \sum_{i=1}^{N}J_{ki}^{2}\,(1-2m_{i})c_{li}+2J_{kl}\,(1-2m_{l})\sum_{i=1}^{N}J_{ki}c_{li}\nonumber \\
 &  & -2J_{kl}^{2}\,c_{ll}(1-2m_{l}),\label{eq:eta_kl_app}
\end{eqnarray}
which yields the expression \prettyref{eq:eta_kl} in the main text.\bigskip{}

\textbf{Fourth order. }The cumulant of fourth order is 
\begin{eqnarray*}
\langle n_{i}n_{j}n_{r}n_{l}\rangle & = & \llangle n_{i}n_{j}n_{r}n_{l}\rrangle\\
 &  & +\llangle n_{i}n_{j}n_{r}\rrangle m_{l}+\llangle n_{j}n_{r}n_{l}\rrangle m_{i}\\
 &  & +\llangle n_{r}n_{l}n_{i}\rrangle m_{j}+\llangle n_{l}n_{i}n_{j}\rrangle m_{r}\\
 &  & +c_{ij}m_{r}m_{l}+c_{ir}m_{j}m_{l}+c_{il}m_{r}m_{j}\\
 &  & +c_{jr}m_{l}m_{i}+c_{jl}m_{r}m_{i}+c_{rl}m_{i}m_{j}\\
 &  & +c_{ij}c_{rl}+c_{ir}c_{jl}+c_{il}c_{rj}\\
 &  & +m_{i}m_{j}m_{r}m_{l},
\end{eqnarray*}
Only those cumulants in which at most two different indices appear
are fixed by first and second order cumulants. We need to distinguish
three cases. The first case is $i=j\neq r=l$ and leads with \eqref{eq:kappa_kkl_kkk}
to the matrix

\begin{eqnarray}
\left\{ \llangle n_{i}n_{i}n_{r}n_{r}\rrangle_{ir}\right\}  & = & \mathbf{C}\circledast(\mathbf{1}-2\mathbf{C})-\diag(\mathbf{C})\,\diag(\mathbf{C})^{T}\nonumber \\
 &  & -2\,\diag(\mathbf{1}-2\mathbf{m}))\,\mathbf{C}\,\diag(\mathbf{m})\nonumber \\
 &  & -2\,\diag(\mathbf{m})\,\mathbf{C}\,\diag(\mathbf{1}-2\mathbf{m})\nonumber \\
 &  & -\diag(\mathbf{C})\left(\mathbf{m}\circledast\mathbf{m}\right)^{T}-\left(\mathbf{m}\circledast\mathbf{m}\right)\diag(\mathbf{C})^{T}\nonumber \\
 &  & -4\,\diag(\mathbf{m})\,\mathbf{C}\,\diag(\mathbf{m})\nonumber \\
 &  & +\mathbf{m}\mathbf{m}^{T}\circledast(\mathbf{1}-\mathbf{m}\mathbf{m}^{T}).\label{eq:ni_ni_nk_nk}
\end{eqnarray}
The second case $i=j=l\neq r$ yields the matrix

\begin{eqnarray}
\left\{ \llangle n_{i}n_{i}n_{i}n_{r}\rrangle_{ir}\right\}  & = & \diag(\mathbf{1}-3\,\diag(\mathbf{C}))\,\mathbf{C}\nonumber \\
 &  & -\diag(\mathbf{C})\mathbf{m}^{T}-3\,\diag(\mathbf{m})\,\mathbf{C}\nonumber \\
 &  & -\left(\mathbf{m}\circledast\diag(\mathbf{C})\right)\,\mathbf{m}^{T}+3\,\diag(\mathbf{m}\circledast\mathbf{m})\,\mathbf{C}\nonumber \\
 &  & +\diag(\mathbf{1}-\mathbf{m}\circledast\mathbf{m}))\,\mathbf{m}\mathbf{m}^{T}\label{eq:ni_ni_ni_nk}
\end{eqnarray}
and the third for $i=j=r=l$ yields a vector

\begin{eqnarray}
\llangle n_{i}n_{i}n_{i}n_{i}\rrangle & = & \mathbf{m}\circledast(\mathbf{1}-(\mathbf{m}\circledast)^{3})-4\,\diag(\mathbf{C})\circledast\mathbf{m}\nonumber \\
 &  & +2\,\diag(\mathbf{C})\circledast(\mathbf{m}\circledast)^{2}\nonumber \\
 &  & -3(\diag(\mathbf{C})\circledast)^{2},\label{eq:ni_ni_ni_ni}
\end{eqnarray}
where we use the notation$\left(\mathbf{x}\circledast\right)^{n}=\mathbf{x}\circledast\ldots\circledast\mathbf{x}$
for the element-wise $n$-th power of a vector. In \prettyref{eq:cond_gain_3rd}
we need the term

\begin{eqnarray*}
\Delta\kappa_{3,kl} & = & \sum_{i,j,r=1}^{N}J_{ki}J_{kj}J_{kr}\llangle n_{i}n_{j}n_{r}n_{l}\rrangle\\
 & \simeq & 3\sum_{i}J_{ki}^{2}J_{kl}\llangle n_{i}n_{i}n_{l}n_{l}\rrangle\qquad{\scriptstyle \left(i=j,r=l\right),\left(i=r,j=l\right),\left(i=l,j=r\right)}\\
 &  & +\sum_{i}J_{ki}^{3}\llangle n_{i}n_{i}n_{i}n_{l}\rrangle\qquad{\scriptstyle \left(i=j=r\neq l\right)}\\
 &  & +3\sum_{i}J_{ki}J_{kl}^{2}\llangle n_{l}n_{l}n_{l}n_{i}\rrangle\qquad{\scriptstyle \left(j=r=l\neq i\right),\left(r=l=i\neq j\right),\left(l=i=j\neq r\right)}\\
 &  & +J_{kl}^{3}\llangle n_{l}n_{l}n_{l}n_{l}\rrangle\qquad{\scriptstyle \left(i=j=r=l\right)},
\end{eqnarray*}
which in matrix form gives rise to expression \prettyref{eq:Delta_kappa3}
of the main text.

\textbf{}

\begin{acknowledgments}
This work was partially supported by the Young investigator's group
VH-NG-1028, Helmholtz Portfolio theme SMHB, and EU Grant 604102 (Human
Brain Project, HBP). All simulations were carried out with NEST (http://www.nest-initiative.org).

\end{acknowledgments}

\end{document}